\documentclass[usegraphicx]{mn2e}
\usepackage[totalwidth=500pt,totalheight=680pt]{geometry}
\usepackage{times}
\newlength{\colwidth}
\newcommand{\ion}[2]{\setcounter{enumiv}{#2}#1\textsc{~\roman{enumiv}}}

\title[Radio imaging of the SXDF - III.]{Radio imaging of the Subaru/XMM-Newton
Deep Field - III. Evolution of the radio luminosity function beyond
$z=1$}
\author[C.\ Simpson et al.]{Chris Simpson$^1$\thanks{E-mail:
cjs@astro.livjm.ac.uk}, Steve Rawlings$^2$, Rob Ivison$^{3,4}$,
Masayuki Akiyama$^5$, \newauthor
Omar Almaini$^6$, 
Emma Bradshaw$^6$, Scott Chapman$^7$, Rob Chuter$^6$, \newauthor
Scott Croom$^8$, Jim Dunlop$^4$, S\'{e}bastien Foucaud$^9$, and
Will Hartley$^6$\\
$^1$Astrophysics Research Institute, Liverpool John Moores University,
Twelve Quays House, Egerton Wharf, Birkenhead CH41~1LD\\
$^2$Department of Physics, University of Oxford, Denys Wilkinson Building,
Keble Road, Oxford OX1~3RH\\
$^3$UK Astronomy Technology Centre, Royal Observatory, Blackford Hill,
Edinburgh EH9~3HJ\\
$^4$Institute for Astronomy, University of Edinburgh, Royal Observatory,
Blackford Hill, Edinburgh EH9~3HJ\\
$^5$Astronomical Institute, Graduate School of Science, Tohoku
University, Aramaki, Aoba, Sendai 980-8578, Japan\\
$^6$School of Physics and Astronomy, University of Nottingham,
University Park, Nottingham NG7~2RD\\
$^7$Institute of Astronomy, University of Cambridge, Madingley Road,
Cambridge CB3~0HA\\
$^8$Sydney Institute for Astronomy, School of Physics, University of
Sydney, NSW~2006, Australia\\
$^9$Department of Earth Sciences, National Taiwan Normal University,
Tingzhou Road, Wenshan District, Taipei 11677, Taiwan}
\begin{document}

\date{Accepted 2012 January 9.  Received 2012 January 9; in original
  form 2011 December 6}

\pagerange{\pageref{firstpage}--\pageref{lastpage}} \pubyear{2012}

\maketitle

\label{firstpage}

\begin{abstract}
We present spectroscopic and eleven-band photometric redshifts for
galaxies in the 100-$\mu$Jy Subaru/\textit{XMM-Newton\/} Deep Field
radio source sample. We find good agreement between our redshift
distribution and that predicted by the SKA Simulated Skies project.
We find no correlation between $K$-band magnitude and radio flux, but
show that sources with 1.4-GHz flux densities below $\sim1$\,mJy are
fainter in the near-infrared than brighter radio sources at the same
redshift, and we discuss the implications of this result for
spectroscopically-incomplete samples where the $K$--$z$ relation has
been used to estimate redshifts. We use the infrared--radio
correlation to separate our sample into radio-loud and radio-quiet
objects and show that only radio-loud hosts have spectral energy
distributions consistent with predominantly old stellar populations,
although the fraction of objects displaying such properties is a
decreasing function of radio luminosity. We calculate the 1.4-GHz
radio luminosity function (RLF) in redshift bins to $z=4$ and find
that the space density of radio sources increases with lookback time
to $z\approx2$, with a more rapid increase for more powerful
sources. We demonstrate that radio-loud and radio-quiet sources of the
same radio luminosity evolve very differently. Radio-quiet sources
display strong evolution to $z\approx2$ while radio-loud AGNs below
the break in the radio luminosity function evolve more modestly and
show hints of a decline in their space density at $z>1$, with this
decline occurring later for lower-luminosity objects. If the radio
luminosities of these sources are a function of their black hole spins
then slowly-rotating black holes must have a plentiful fuel supply for
longer, perhaps because they have yet to encounter the major merger
that will spin them up and use the remaining gas in a major burst of
star formation.
\end{abstract}


\section{Introduction}

Deep radio surveys provide an excellent means for studying two of the
most important processes in galaxy evolution: star formation, and
accretion onto supermassive black holes. These two processes are
believed to be linked by a mechanism or mechanisms known as
`AGN-driven feedback' (e.g., Croton et al.\ 2006; Bower et al.\ 2006)
that produces the observed correlation between the masses of the black
hole and stellar bulge (e.g., Ferrarese \& Merritt 2000; Gebhardt et
al.\ 2000) and therefore radio surveys can provide insight into this
aspect of galaxy evolution.

At 1.4-GHz flux densities brighter than a few mJy, the AGN population
is dominated by `radio-loud' sources, whose powerful radio-emitting
jets carry a kinetic power comparable to their optical/UV/X-ray
photoionizing luminosity (e.g., Rawlings \& Saunders 1991). This
population has been further subdivided in two different ways. Fanaroff
\& Riley (1974) noticed a dichotomy in radio morphologies between
`edge-darkened' sources where the low brightness regions are further
from the galaxy than the high brightness regions, and
`edge-brightened' sources where the opposite is true. Fanaroff \&
Riley noted that the relative numbers of these two classes (now called
Fanaroff \& Riley Class~I and Class~II respectively) changed sharply
at a 178-MHz luminosity of $L_{\rm178MHz} \approx
10^{25}\rm\,W\,Hz^{-1}$ (corrected from their assumed value of
$H_0=50\rm\,km\,s^{-1}\,Mpc^{-1}$ to
$H_0=72\rm\,km\,s^{-1}\,Mpc^{-1}$). The physical reason for this
morphological dichotomy is understood to be the speed of the jets,
with the jets of FR\,I sources having low Mach numbers and being
susceptible to turbulence, while FR\,II jets remain supersonic beyond
the confines of the galaxy until they terminate in shocks at the radio
hotspots (Bicknell 1985). This is supported by the apparent variation
in the radio luminosity of the Fanaroff--Riley break with host galaxy
optical luminosity (Owen \& White 1991; Ledlow \& Owen 1996),
indicating that jets are more rapidly decelerated in more massive host
galaxies (De Young 1993).

The dependence on the kpc-scale host galaxy luminosity indicates that
the Fanaroff \& Riley class is likely to be extrinsic to the accreting
supermassive black hole which powers these objects. A more fundamental
dichotomy is seen in the optical spectra of radio galaxies (Hine \&
Longair 1979; Laing et al.\ 1994). Hine \& Longair subdivided objects
as Class~A or Class~B, with Class~A sources displaying rich
emission-line spectra and Class~B objects showing no line emission or
only weak emission from low-ionization species. They discovered a
strong correlation between radio luminosity and the fraction of
Class~A sources, and it is now believed that this split is due to
differences close to the black hole, with Class~B sources possessing a
dearth of ionizing photons due to radiatively-inefficient accretion
which does not proceed through a disc. According to Hine \& Longair
(1979), the majority of radio sources with
$L_{\rm178MHz}\ga10^{26.2}\rm\,W\,Hz^{-1}$ (again, corrected for their
assumed value of the Hubble constant) have Class~A spectra.  This
luminosity is above the Fanaroff \& Riley break and is close to the
break in the radio luminosity function (e.g., Dunlop \& Peacock 1990;
Willott et al.\ 2001).  In the unified model for active galactic
nuclei (e.g., Antonucci 1993), Class~A sources should also display an
infrared excess as the ionizing photons heat the torus and are
reprocessed as thermal radiation. Vardoulaki et al.\ (2008, hereafter
Paper~II) used \textit{Spitzer\/}/MIPS 24-$\mu$m imaging to classify
radio sources in the Subaru/\textit{XMM-Newton\/} Deep Field (SXDF)
and demonstrated that the Class~A/B dichotomy persists to $z\ga1$.

It has long been known that the population of the most powerful radio
sources evolves more strongly than the less powerful population
(Dunlop \& Peacock 1990) and this is demonstrated in differing
evolution between the FR\,I and FR\,II populations. However, we follow
Willott et al.\ (2001) in believing the Hine \& Longair classification
to be the more fundamental division of the radio-loud AGN
population. The weaker evolution seen in the FR\,I population (e.g.,
Clewley \& Jarvis 2004; McApline \& Jarvis 2011) is therefore a
consequence of these objects being exclusively low-luminosity
sources. This view is supported by recent studies which show that
FR\,I and FR\,II sources of similar luminosities display similar
evolution (Rigby, Best \& Snellen 2008), and suggests that a complete
understanding of radio source evolution can only be obtained with the
benefit of significant complementary data to aid in the classification
of sources.

In Simpson et al.\ (2006; hereafter Paper~I), we presented a catalogue
of 505 sources with 1.4-GHz peak radio flux densities greater than
100\,$\mu$Jy over a 0.81-square degree region of the SXDF and their
optical counterparts. By studying the nature of these optical
counterparts as a function of radio flux density, we presented the
first observational evidence for a significant contribution to the
radio source counts below $\sim300\,\mu$Jy from `radio-quiet' AGNs, by
which we mean supermassive black holes with a high accretion rate but
which lie well below the correlation of Rawlings \& Saunders
(1991). This population exists in addition to the low-luminosity tail
of the radio-loud AGN population and the population of star-forming
galaxies which had previously been identified at these fluxes. Our
result was at odds with some previous syntheses of the radio source
counts (e.g., Seymour, McHardy \& Gunn 2004) but in agreement with
others (Jarvis \& Rawlings 2004). Subsequent studies in other deep
survey fields by Seymour et al.\ (2008), Smol\v{c}i\'{c} et
al.\ (2008), and Ibar et al.\ (2009) have since confirmed the presence
of many radio-quiet AGNs at these radio flux densities. Since radio
emission is not obscured or attenuated by dust or gas, deep radio
surveys offer a unique opportunity to identify heavily-absorbed AGNs
which would be missed at other wavelengths. Mart\'{\i}nez-Sansigre et
al.\ (2007) have already used the radio data of Paper~I to suggest
that Compton-thick AGNs may outnumber quasars at high redshift.

To address these issues, we are continuing our detailed investigation
of radio sources in the SXDF. Possessing excellent optical and
near-infrared data over nearly one square degree, this is an ideal
field in which to study the properties of radio sources, especially at
$z\ga1$. The format of this paper is as follows. In Section~2 we
describe new and existing spectroscopic observations of the
100-$\mu$Jy SXDF sample of Paper~I and the manner in which redshifts
were determined. Section~3 describes the imaging data and derivation
of photometric redshifts for the sample. In Section~4 we use these
redshifts to study the $K$-band Hubble diagram at faint radio fluxes
and investigate the relationship between radio-loudness and the
stellar population of the host galaxy. We also determine the cosmic
evolution of the entire radio source population and the radio-loud and
radio-quiet sources separately. Finally, in Section~5, we summarize
our main results. Throughout this paper we adopt a $\Lambda$CDM
cosmology with $\Omega_\Lambda = 1 - \Omega_{\rm m} = 0.74$ and $H_0 =
72\rm\,km\,s^{-1}\,Mpc^{-1}$, as derived from the five-year
\textit{Wilkinson Microwave Anisotropy Probe\/} data (Dunkley et
al.\ 2009). In converting between radio flux (or luminosity) densities
at different frequencies we assume a power-law spectrum of the form
$S_\nu \propto \nu^{-0.7}$, where $S_\nu$ is the flux density at a
frequency $\nu$.

\section{Spectroscopic observations}

\subsection{VIMOS spectroscopy}

\subsubsection{Target selection and observation strategy}

Many of the spectra presented here were obtained as part of the
European Southern Observatory (ESO) programme P074.A-0333, undertaken
using the Visible Multi-Object Spectrograph (VIMOS) instrument on
UT3/Melipal. VIMOS (Le F\`{e}vre et al.\ 2003) consists of four
separate quadrants, each $7'\times8'$, separated by gaps approximately
$2'$ wide, which can be used in imaging or multi-object spectroscopy
modes.  The primary science goal of this programme was to study the
accretion history of the Universe by measuring redshifts for radio and
X-ray-selected galaxies. Targets were prepared from preliminary
optical identification lists for radio and X-ray sources (the latter
provided by M.~Akiyama, private communication). Our observations were
made in service mode using the MR-Orange grating and the GG475
order-sorting filter, which provides a spectral resolution of
$\lambda/\Delta\lambda\approx580$ (1-arcsecond slit) over the
wavelength range $4800{\rm\AA} < \lambda < 10\,000{\rm\AA}$, with a
sampling of 2.5\,\AA\,pixel$^{-1}$. Twenty-seven slit masks were
prepared, with the first being positioned where it provided the
maximum number of radio and X-ray source spectra, accounting for
overlap in the dispersion direction. The objects that were selected
for the first mask were then removed from the target catalogue and the
optimum location for the second mask was determined, and so on for the
third and subsequent masks. The 27 masks designed in this manner
provided spectroscopic observations for $\sim80$\% of the radio and
X-ray sources.

To prepare the slit masks, five minute $R$-band images at each mask
position were obtained and the VIMOS Mask Preparation Software (VMMPS)
was used to provide the focal plane distortion corrections. The
$R$-band catalogues of Furusawa et al.\ (2008) provided the celestial
coordinates of objects detected in the pre-imaging, from which a
geometric mapping between celestial and focal plane coordinates was
calculated. After assigning slits using only a list of radio and X-ray
sources as the input catalogue, additional slits were assigned to
sources from a much larger list consisting of lower-significance X-ray
sources, submm-detected objects from the SHADES survey (Coppin et
al.\ 2006), Ly$\alpha$ emitters (Ouchi et al.\ 2008), and Lyman break
galaxies.

Each field was observed as two separate observation blocks (OBs), one
consisting of a single 2700-second exposure, and the other comprising
two 1350-second exposures. Although the fields were assigned
priorities according to the order in which their positions were
determined (and hence in order of decreasing numbers of primary
targets), these priorities were initially ignored by the service
observers in favour of a strategy of observing the field with the
lowest airmass. Since the programme was terminated after two years
when only $\sim75$\% complete, this strategy resulted in fewer spectra
being obtained.

\subsubsection{Data reduction}

The spectra were reduced within the framework of EsoRex Version~3.5.1,
although initial reduction attempts revealed numerous problems with
the pipeline recipes which had to be overcome.

The individual spectra in each exposure were identified, traced, and
distortion corrections calculated by the recipe \textit{vmspcaldisp\/}
using daytime arc lamp exposures, with these corrections then being
applied to the science frames. This recipe should require no user
input but it failed to produce suitable results in all but a few
cases. After discussion with the ESO User Support Department, we were
able to obtain acceptable results by running the recipe many times
with different values of the FITS header keyword \texttt{ESO PRO OPT
  DIS X\_0\_0} and choosing the reduced frame which produced the lowest
rms deviation in the fit to the wavelengths of the arc lines.

The recipe \textit{vmmosobsstare\/} uses the output from
\textit{vmspcaldisp\/} to extract the two-dimensional slit spectra
from individual science exposures and resample them to a common linear
wavelength solution. The data were also flat-fielded and cleaned of bad
pixels during this process. In addition to an image of aligned
spectra, this recipe also produced a table listing the image rows
occupied by each slit spectrum and the locations of objects detected
in the slits.

Later reduction steps were performed within the \textsc{iraf}
package.\footnote{\textsc{iraf} is distributed by the National Optical
  Astronomy Observatories, which are operated by the Association of
  Universities for Research in Astronomy, Inc., under cooperative
  agreement with the National Science Foundation.} Background
subtraction was generally undertaken by cubic splines in the spatial
direction, masking out regions where objects had been identified and
applying a sigma-clipping algorithm. By default, the top and bottom
rows of each slit spectrum were not included in the background
determination, but occasionally more rows had to be excluded. The
number of spline pieces was chosen to ensure that splines did not
connect within a masked region, and that each spline covered at least
30 pixels. For Ly$\alpha$ emitters (for which results were presented
in Ouchi et al.\ 2008), a simple median sky subtraction was applied at
each wavelength position.

The recipe \textit{vmmosstandard\/} was used to reduce the
spectrophotometric standard star observations. This produces a
response curve by comparing the observed count rate to the known flux
density spectrum of the standard. The standard pipeline reduction then
median-filters and heavily smooths this response curve before it is
applied to the object spectra. However, these steps produce data of
unacceptable quality since they wash out the relatively high frequency
response variations caused by the order sorting filter at the blue
end, as well as failing to account for contamination from higher
orders at the red end (the spectrophotometric standards are usually
blue, so this is a major problem for the standard stars but irrelevant
for the science targets). Instead, a hybrid response curve was applied
to the data. At wavelengths $\lambda>8500$\,\AA, a quadratic
extrapolation of the smoothed response curve was applied, while at
$\lambda\la6500$\,\AA\ the raw response curve was used -- the
signal-to-noise ratio of the standard star spectra is far higher than
any of the object spectra, so this does not increase the noise by any
measurable amount. Between these wavelengths, the smoothed response
curve was used.

Visual inspection of the reduced images from a single mask taken with
different OBs revealed that there were sometimes quite significant
spatial shifts and non-uniform flux variations which were believed to
be due to instrument flexure and/or mask misalignment. Additional
problems were caused by some spectra being taken in inappropriate
conditions such as through heavy cloud or in poor seeing. This
prevented a simple coaddition of the three individual two-dimensional
spectra. Instead, one-dimensional spectra were extracted (typically
using a 1-arcsecond wide aperture) and combined using a sigma-clipping
algorithm with the noise level estimated from the pixel-to-pixel
variations in the spectrum. For objects where the continuum was
detected at a high signal-to-noise ratio, the spectra were first
normalized to the flux density level of the brightest spectrum before
combining. From adding artificial emission lines to the reduced
spectra, we estimate that the limiting line flux over the most
sensitive part of the spectrum ($5300{\rm\,\AA} \la \lambda \la
7700\rm\,\AA$) is $\sim10^{-20}$\,W\,m$^{-2}$, although the issues
described above mean that an intrinsically brighter line could fail to
be detected.

\subsection{Other spectroscopic data}
\label{sec:otherspec}

Several observational campaigns have obtained spectra of objects
within the SXDF/UDS, and Paper~II presented spectra for 28 of the
brightest 37 radio sources, obtained from a variety of sources.
Sometimes these observations targeted radio sources with spare slits
or fibres that could not be assigned to primary targets but we also
cross-correlated the radio catalogue with the targets of all known
spectroscopic observations in order to obtain as many redshifts as
possible. We briefly describe the origins of our spectra here.

Radio sources were targeted by three campaigns. Geach et al.\ (2007)
took spectra of galaxies in intermediate-redshift groups and clusters
containing low-power radio sources with the LDSS2 spectrograph on the
Magellan telescope. Smail et al.\ (2008) observed candidate
KX-selected quasars (Warren, Hewett \& Foltz 2000) with the AAOmega
spectrograph while Akiyama et al.\ (in preparation) followed up X-ray
sources from the catalogue of Ueda et al.\ (2008) using Subaru
Telescope's FOCAS spectrograph. In these latter two cases spare fibres
or slits were allocated to objects in the 100-$\mu$Jy sample, while
some X-ray targets are also radio sources and so where primary targets
of the FOCAS observations.

Akiyama (private communication) used the 2-degree field spectrograph
(2dF) to observe all optically-bright sources in the SXDF and this has
provided many spectra in this paper. A small number of radio sources
were observed with Keck/DEIMOS by van Breukelen et al.\ (2009) in
their sample of colour-selected clusters at $z\sim1$, and by Banerji
et al.\ (2011) in their study of luminous star-forming galaxies. There
is also overlap between the 100-$\mu$Jy sample and the
\textit{Spitzer\/} sources observed with the AAOmega spectrograph in
October 2006 by Scott Croom (private communication). The 580V and 385R
gratings were used to provide a resolution of $R\approx1300$ over the
wavelength range 3700--8800\,\AA\ and the total exposure time was
90\,minutes.

The UDSz European Southern Observatory Large Program (Program ID
180.A-0776(B); PI O.~Almaini) has taken $\sim$3600 spectra of galaxies
in the UDS field. The majority of these were selected to lie at $z>1$
based on photometric redshifts, and several are counterparts to
sources in the 100-$\mu$Jy catalogue.  Spectra were taken with either
VIMOS or FORS2, depending on the magnitudes and colours of the
targets. The VIMOS data reduction is described in Chuter (2011) and
Almaini et al.\ (in preparation) while a description of the FORS2 data
reduction will appear in Pearce et al.\ (in preparation).

Finally, the spectrum of VLA~0054 was obtained using the ISIS
double-arm spectrograph on the 4.2-m William Herschel Telescope on the
night of UT 2010 Feb 12. The set-up and reduction method were as
described in Jarvis et al.\ (2009).

In some cases the same source was observed multiple times with
different instruments and, in such cases, we have confirmed that the
redshifts (where the data were of sufficiently high quality to obtain
one) agree. In deciding which spectrum to use for the redshift
determination, we generally choose the spectrum with the highest
signal-to-noise ratio although sometimes a lower-quality spectrum with
greater wavelength coverage may include additional spectral features
that improve the result. In the Appendix we plot the spectra of all
objects for which we have derived redshifts, plus additional VIMOS
spectra where only featureless continuum was detected.

\subsection{Redshift determination}

All spectroscopic redshifts presented here were calculated using the
\textsc{iraf} task \textit{fxcor\/}, which performs a Fourier cross
correlation between the target spectrum and a template spectrum and
applies a correction to obtain a heliocentric recession velocity. Most
of the templates used were from the Sloan Digital Sky Survey (SDSS),
after having their wavelength scales converted from vacuum to air
(I.~Baldry, private communication). A pure emission-line template was
also constructed, consisting of the lines in Ferland \& Osterbrock
(1986) and McCarthy (1993), normalized by their [\ion{O}{3}]
fluxes. To provide templates for higher-redshift galaxies, spectra
were produced from the Bruzual \& Charlot (2003) spectral synthesis
code: a `starburst', consisting of a 290-Myr-old solar metallicity
population with an [\ion{O}{2}] emission line added; and a `radio
galaxy', comprising a 1.4-Gyr-old solar metallicity stellar population
with the emission line template added. These spectra were
cross-correlated with the SDSS templates and the velocity differences
were $<10\,$km\,s$^{-1}$.

Typical redshift uncertainties for objects for which redshifts were
measured from narrow emission and/or absorption features were
estimated from the width of the cross-correlation peak, and found to
be of the order of 50--100\,km\,s$^{-1}$. A similar level of
uncertainty was found to arise from the choice of wavelength range
over which the cross-correlation was performed, which sometimes had to
be selected to eliminate spurious features resulting from cosmic rays
or the effects of fringing. The uncertainties for sources with broad
emission lines are significantly larger.

In cases where no redshift was readily determined, the reduced
two-dimensional spectra were inspected to search for possible emission
lines, especially in the region $\lambda \ga 7700$\,\AA\ which is
strongly affected by fringing. Where there were concerns that features
may have been artifacts of the data reduction, it was confirmed that
they existed in the raw frames.

Spectral classifications were assigned based on observed emission
and/or absorption line properties. Objects with broad permitted lines
were classified as broad-line AGNs (BLAGN), while those displaying no
emission lines were classified as absorption-line galaxies (Abs). For
objects displaying a wider spectrum of narrow emission lines, the
criterion of Simpson (2005) using the [\ion{N}{2}]/H$\alpha$ vs
[\ion{O}{3}]/H$\beta$ diagram of Baldwin, Phillips \& Terlevich (1981)
was used wherever possible. Alternatively, the presence of strong,
high-ionisation ultraviolet emission lines such as \ion{C}{4} and
[\ion{Ne}{3}] were used to diagnose the presence of a narrow-line AGN
(NLAGN). For intermediate-redshift objects where the wavelength
coverage did not include H$\alpha$, a classification was made based on
another line-ratio diagram from Baldwin et al.\ (1981). For many
objects, [\ion{O}{2}] was the only strong emission line visible and, in
these cases, the presence of strong \ion{Mg}{2} absorption was
considered to indicate a star-forming galaxy (e.g., Kinney et
al.\ 1996). Finally, in uncertain cases, an object was simply
classified as a `weak' or `strong' line-emitter according to whether
the rest-frame equivalent width of the [\ion{O}{2}] line was less or
more than 15\,\AA, respectively.

In total, we have derived redshifts for 267 of the 505 sources in the
100-$\mu$Jy catalogue, of which 256 are robust.

\section{Photometric redshifts}

\subsection{Imaging data}

\subsubsection{Infrared photometry}
\label{sec:IRphotom}

The near-infrared data used here come from the Third Data Release
(DR3) of the UKIRT Infrared Deep Sky Survey (UKIDSS; Lawrence et
al.\ 2007). The Ultra Deep Survey (UDS) covers an area of 0.77 square
degrees centred at $\alpha$=02$^{\rm h}$17$^{\rm m}$48$^{\rm s}$
$\delta$=$-$05$^\circ$06$'$00$''$ (J2000.0) and provides approximately
0.63 square degrees of overlap with the Subaru/\textit{XMM-Newton\/}
Deep Field. The UDS DR3 data reach 5-$\sigma$ depths of $J=22.8$
(23.7), $H=22.1$ (23.5), and $K=21.8$ (23.7) in Vega (AB). Sources
from the $K$-band-selected catalogue of S.~Foucaud (private
communication) were selected as infrared counterparts to the radio
sources by visual matching to the optical counterparts from
Paper~I. In instances where there was no near-infrared counterpart in
the DR3 catalogue, the flux was measured directly from the images,
either at the radio position or the location of the optical
counterpart from Paper~I (after applying the astrometric correction
described below).

We repeated our astrometric comparison, this time between the radio
and near-infrared frames, and find an almost identical offset to that
between the radio and optical images from Paper~I, with the $K$-band
sources lying, on average, 0.05\,arcsec west and 0.22\,arcsec south of
the radio positions. This consistency is not unexpected, since both
the optical and near-infrared astrometric solutions were based on the
Two Micron All-Sky Survey (2MASS; Skrutskie et al.\ 2006). We note
that Morrison et al.\ (2010) have proposed that astrometric offsets
such as this might be caused by the VLA on-line system although, since
they are much smaller than either the optical or radio point spread
functions, we have not investigated whether this is indeed the cause.

All measurements were made in a 3-arcsec diameter aperture. Although
this provides a lower signal-to-noise ratio measurement and is more
susceptible to source blending than a smaller aperture (e.g., the
1.75-arcsec aperture used by Williams et al.\ 2009), colours derived
in this aperture are far less sensitive to mismatched point spread
functions and astrometric registration errors. For those sources which
are not within the UDS footprint, photometric measurements were made
from the 2MASS images, using the calibration information given in the
image headers. Due to the larger point spread function and coarser
pixel scale of the 2MASS images, measurements were made in 4-arcsec
apertures. A combined aperture and filter correction was determined by
comparing the UDS and 2MASS photometry for all galaxies within the
UDS: an offset of $m_{\rm2MASS}(\phi4'')-m_{\rm UDS}(\phi3'')=0.20$
was found for all filters over the range $14<K_{\rm UDS}<17$ and
applied to the measurements. Although the 2MASS data are $\sim5$\,mag
shallower than the UDS images, the addition of even shallow
near-infrared photometry was found to substantially improve the
quality of the photometric redshifts (see below). All objects outside
the UDS footprint had their 2MASS images inspected and the 2MASS
photometry was not used when it was contaminated by a nearby star or
galaxy.

Additional, longer-wavelength measurements were made from the
\textit{Spitzer\/}/IRAC images released by the \textit{Spitzer\/}
Wide-Area Infrared Extragalactic Survey (SWIRE; Lonsdale et
al.\ 2003). Only the IRAC1 (3.6\,$\mu$m) and IRAC2 (4.5\,$\mu$m) data
were used. Again, a correction had to be applied due to the much
larger point spread function (2.1$''$ FWHM) of the IRAC data. We
assumed that the morphologies of objects in the IRAC filters followed
those in the $K$-band image, and smoothed the $K$ image to match the
resolution of the IRAC data. Photometry was then measured from both
images in 3-arcsecond apertures for all sources in the DR3 catalogue,
and the following relationship was found to provide a good fit to the
mean of the data:
\begin{equation}
K_{\rm smooth} - K_{\rm raw} = 0.27 + 0.01(K_{\rm raw}-20) 
\label{eqn:psfcorr}
\end{equation}
where $K_{\rm smooth}$ and $K_{\rm raw}$ are the Vega magnitudes measured
in the smoothed and unsmoothed $K$-band images, respectively. This
correction factor was applied to the measured IRAC aperture photometry
of sources outside the UDS field of view, while the
individually-determined correction factors were applied to UDS
sources. In both cases, an additional photometric uncertainty of
0.03\,mag was added to the IRAC photometry, corresponding to the
r.m.s.\ of the distribution about the mean given in
Equation~\ref{eqn:psfcorr}.

\subsubsection{Optical photometry}

The optical data in the UDS come from the Subaru/\textit{XMM-Newton\/}
Deep Field, which comprises five separate Suprime-Cam pointings
(Furusawa et al.\ 2008). To extract photometric information from these
images, residual background variations were removed by subtracting a
low spatial frequency background image computed by SExtractor (Bertin
\& Arnouts 1996). An astrometric distortion map with respect to the
UDS astrometric frame was calculated for each SXDF image as
follows. All UDS sources were matched to each of the $i'$-band source
catalogues, using a 1-arcsecond matching radius. The distortion at the
position of each UDS source was then estimated as the mean offset
between UDS and SXDF positions for all sources within
1\,arcminute. Separate polynomial fits to the right ascension and
declination distortions were then calculated, using the lowest orders
which produced an r.m.s.\ deviation of $<25$\,milliarcseconds. In all
cases, the maximum deviation was less than 125\,mas in each
coordinate.

Photometry was measured for each UDS source in a 3-arcsecond diameter
aperture at the distortion-correction position within each SXDF
image. In cases where more than one Suprime-Cam image covered the
position of the source, preference was given to an image where the
measurement apertures (which extend up to a diameter of 5\,arcseconds)
did not include any masked region (i.e., the cores of bright stars and
the CCD bleeds). If the photometry from two images was clean,
preference was given to photometry measured from the Suprime-Cam image
on which the source was furthest from the edges. If three images
produced clean photometry, one set was removed from consideration if
its flux was more than ten times further from the median than the
other extreme datapoint; the image on which the source was furthest
from the edges was then used. No corrections are made for differences
in the point spread functions of the Subaru and UKIRT images, since
the FWHM of each image is always measured to be within the range
0.77--0.84\,arcsec.

In seven cases, the radio sources lie outside the footprint of the
optical data from the SXDF. However, Subaru/Suprime-Cam \textit{RI\/}
imaging exists over this area (M.~Akiyama, private communication) and
photometry was measured in 3-arcsec apertures for these sources.

A $u^*$-band image covering the entire 100-$\mu$Jy catalogue footprint
was provided by S.~Foucaud and is composed of data from the
Canada--France--Hawaii Telescope Legacy Survey (CFHTLS) plus
additional imaging taken in support of the UDS. It has poorer image
quality than the SXDF and UDS imaging (1.0-arcsec FWHM) so a correction
has been applied to the 3-arcsec aperture magnitudes. This was
determined by smoothing the $B$-band image to the same point spread
function and comparing the aperture magnitudes in the smoothed and
unsmoothed images. A difference of 0.025\,mag was calculated, and this
was applied to all $u^*$ fluxes before photometric redshifts were
calculated. Although we apply a single correction to all objects,
there is no apparent variation in its size with magnitude, nor does
the object-to-object variation exceed other sources of noise; e.g., at
$u^*=25.0$ the standard deviation of the magnitude differences between
the smoothed and unsmoothed images is 0.09\,mag but the measurements
themselves only have a signal-to-noise ratio of $\sim8$.

We estimate the photometric uncertainty in each optical filter by
fitting a Gaussian to a histogram of the counts measured in 5000
randomly-placed apertures on each image and using the standard
deviation of this Gaussian as the 1-$\sigma$ error. Each Suprime-Cam
image was split into an 8$\times$10 grid (each grid element comprising
approximately 1000$^2$ pixels) before undertaking this procedure. We
also investigated the variation of noise with aperture radius $r$ and
found a dependence approximately given by $r^{1.75}$. This is
plausible for deep images where confusion noise is important (such as
the SXDF images), as the noise should vary $\propto r$ (i.e., in the
same manner as background noise) for steep number counts and $\propto
r^2$ for flat counts (Condon 1974), and galaxy number counts flatten
near the SXDF limiting magnitudes (Furusawa et al.\ 2008, and
references therein).

\subsection{Sources excluded from the photometric redshift
  analysis}
\label{sec:exclusion}

Three sources (VLA0091, VLA0133 and VLA0462) lie too close to bright
stars for reliable photometry to be obtained and have been excluded
without biasing any results. Two further radio sources (VLA0425 and
VLA0492) lie in the outskirts of bright galaxies with photometric
redshifts of 0.40 and 0.09, respectively, and no counterparts are seen
at the location of the radio sources. The latter of these could
plausibly be a supernova remnant (its 1.4-GHz luminosity is
$2.1\times10^{21}$\,W\,Hz$^{-1}$), but the former is $\sim20$ times
more luminous and is more likely to be a background source. Both
objects are excluded from further analysis without biasing the
results. VLA0081 and VLA0171 are excluded from the analysis because,
although they have visible counterparts, they are too faint and too
close to nearby stars for accurate photometric measurements. As we
might have been able to obtain reliable photometry had these sources
been brighter in the optical/infrared, their exclusion could bias our
results. However, there are no other objects in the catalogue that are
this close to stars of similar brightness and therefore they can be
excluded solely on the basis of their proximity to bright sources, and
not the faintness of their counterparts.

All objects which lacked optical counterparts in Paper~I now have
plausible detections in the UKIDSS UDS and/or IRAC images, permitting
the derivation of photometric redshifts.

\subsection{Derivation of photometric redshifts}

\begin{table*}
\caption[]{Photometric and (where available) spectroscopic redshifts of all 505 SXDF radio sources. The probability-weighted mean photometric redshift is given, together with its 68\,per cent confidence interval. A `?' in the $z_{\rm spec}$ column indicates a spectrum was taken with VIMOS but no plausible redshift could be determined. The column headed `Type' presents the source classification, described in Section~2.3, while a `Cont' in this column indicates the detection of continuum emission and the spectrum is therefore presented in Fig.~A1. References for the spectroscopy are as follows: (1) This work; (2) Paper~II; (3) Geach et al.\ (2007); (4) van Breukelen et al.\ (2009); (5) Smail et al.\ (2008); (6) Akiyama et al.\ (in prep); (7) S.~Chapman (priv.\ comm.); (8) S.~Croom (priv.\ comm.); (9) UDSz/VIMOS (Chuter 2011); (10) UDSz/FORS2 (Pearce et al., in prep).\label{tab:data}}
\begin{tabular}{lccccc@{}llcl}
\hline
Name & \multicolumn{2}{c}{IR position (J2000.0)}
& $K$ ($\phi4''$) & $z_{\rm phot}$ & \multicolumn{2}{c}{$z_{\rm spec}$}
 & Type & Ref & Comments \\ \hline
VLA0001 & 02:18:27.19 & $-$04:54:41.2 & 16.58$\pm$0.00 & 0.58 (0.55--0.61) &  0.6272 &   & NLAGN& 1 &  \\
VLA0002 & 02:18:18.14 & $-$04:46:07.2 & 17.60$\pm$0.00 & 1.00 (0.95--1.04) &  1.1821 &   & NLAGN& 2 &  \\
VLA0003 & 02:18:39.55 & $-$04:41:49.4 & 19.03$\pm$0.02 & 2.18 (2.13--2.23) &  2.4349 &   & NLAGN& 1 &  \\
VLA0004 & 02:18:53.63 & $-$04:47:35.6 & 19.51$\pm$0.02 & 2.57 (2.42--2.72) &  2.4719 &   & NLAGN?& 1 &  \\
VLA0005 & 02:18:51.38 & $-$05:09:01.6 & 18.75$\pm$0.01 & 1.75 (1.65--1.85) & ? &   && 1 &  \\
VLA0006 & 02:16:37.82 & $-$05:15:28.3 & 18.12$\pm$0.01 & 1.55 (1.41--1.68) &  &   &&  & {\it uBVRiz\/} affected by stellar halo \\
VLA0007 & 02:16:59.03 & $-$04:49:20.3 & 17.79$\pm$0.01 & 1.31 (1.26--1.36) &  1.3247 &   & NLAGN& 6 &  \\
\hline
\end{tabular}
\end{table*}
\addtocounter{table}{-1}

\begin{table*}
\caption[]{\textit{continued}.}\begin{tabular}{lccccc@{}llcl}
\hline
Name & \multicolumn{2}{c}{IR position (J2000.0)}
& $K$ ($\phi4''$) & $z_{\rm phot}$ & \multicolumn{2}{c}{$z_{\rm spec}$}
 & Type & Ref & Comments \\ \hline
VLA0060 & 02:18:22.30 & $-$05:29:09.1 & 17.33$\pm$0.00 & 0.82 (0.79--0.86) &  0.8805 &   & Weak& 1 &  \\
VLA0061 & 02:17:02.68 & $-$05:16:18.5 & 18.73$\pm$0.01 & 0.64 (0.59--0.68) &  0.6682 &   & Strong& 9 &  \\
VLA0062 & 02:17:52.03 & $-$05:03:08.1 & 19.82$\pm$0.03 & 2.71 (2.58--2.85) & ? &   && 1 &  \\
VLA0063 & 02:19:36.94 & $-$05:13:42.5 &       $>$20.93 & 1.93 (1.26--2.64) &  &   &&  &  \\
VLA0064 & 02:18:49.02 & $-$04:59:31.6 & 16.86$\pm$0.00 & 0.51 (0.48--0.54) &  0.5152 &   & Strong& 1 &  \\
VLA0065 & 02:18:42.05 & $-$05:32:50.8 & 16.35$\pm$0.00 & 0.39 (0.37--0.41) &  0.3834 &   & Strong& 3 &  \\
VLA0066 & 02:17:32.19 & $-$04:31:10.6 &       $>$15.57 & 1.04 (1.01--1.07) &  &   &&  &  \\
\hline
\end{tabular}
\end{table*}
\addtocounter{table}{-1}

\begin{table*}
\caption[]{\textit{continued}.}\begin{tabular}{lccccc@{}llcl}
\hline
Name & \multicolumn{2}{c}{IR position (J2000.0)}
& $K$ ($\phi4''$) & $z_{\rm phot}$ & \multicolumn{2}{c}{$z_{\rm spec}$}
 & Type & Ref & Comments \\ \hline
VLA0124 & 02:17:04.77 & $-$05:15:18.1 & 17.59$\pm$0.00 & 1.26 (1.21--1.31) &  1.2793 &   & Abs& 10 &  \\
VLA0125 & 02:19:11.08 & $-$04:56:33.6 & 20.01$\pm$0.04 & 1.69 (1.58--1.79) & ? &   && 1 &  \\
VLA0126 & 02:16:40.85 & $-$05:15:30.9 & 20.96$\pm$0.10 & 2.44 (2.18--2.66) &  &   &&  &  \\
VLA0127 & 02:18:30.69 & $-$05:00:55.0 & 18.00$\pm$0.01 & 0.87 (0.82--0.91) &  0.8797 & ? & SB& 1 &  \\
VLA0128 & 02:18:06.24 & $-$04:55:38.1 & 20.83$\pm$0.08 & 1.93 (1.36--2.63) &  &   &&  &  \\
VLA0129 & 02:18:31.38 & $-$05:36:31.3 &       $>$15.88 & 1.35 (1.33--1.43) & ? &   & Cont& 1 &  \\
VLA0130 & 02:17:09.36 & $-$04:37:09.2 & 15.86$\pm$0.00 & 0.39 (0.36--0.41) &  0.3739 &   & Weak& 6 &  \\
\hline
\end{tabular}
\end{table*}
\addtocounter{table}{-1}

\begin{table*}
\caption[]{\textit{continued}.}\begin{tabular}{lccccc@{}llcl}
\hline
Name & \multicolumn{2}{c}{IR position (J2000.0)}
& $K$ ($\phi4''$) & $z_{\rm phot}$ & \multicolumn{2}{c}{$z_{\rm spec}$}
 & Type & Ref & Comments \\ \hline
VLA0188 & 02:17:44.09 & $-$05:28:09.8 & 18.20$\pm$0.01 & 1.18 (1.10--1.27) & ? &   && 1 &  \\
VLA0189 & 02:18:59.92 & $-$04:41:26.4 & 18.35$\pm$0.01 & 0.45 (0.42--0.47) &  &   &&  &  \\
VLA0190 & 02:17:43.03 & $-$04:36:25.0 & 16.25$\pm$0.00 & 0.22 (0.20--0.25) &  0.7847 &   & BLAGN& 6 &  \\
VLA0191 & 02:17:22.04 & $-$04:58:53.1 & 17.98$\pm$0.01 & 0.98 (0.94--1.03) &  1.0960 &   & NLAGN& 6 &  \\
VLA0192 & 02:18:02.07 & $-$04:29:12.6 &       $>$15.84 & 1.64 (1.52--1.75) & ? &   &&  &  \\
VLA0193 & 02:17:17.11 & $-$05:33:32.1 &       $>$15.77 & 1.26 (1.15--1.36) & ? &   &&  &  \\
VLA0194 & 02:19:03.49 & $-$04:39:35.2 & 17.63$\pm$0.01 & 1.08 (1.06--1.14) &  1.8385 &   & BLAGN& 8 &  \\
\hline
\end{tabular}
\end{table*}
\addtocounter{table}{-1}

\begin{table*}
\caption[]{\textit{continued}.}\begin{tabular}{lccccc@{}llcl}
\hline
Name & \multicolumn{2}{c}{IR position (J2000.0)}
& $K$ ($\phi4''$) & $z_{\rm phot}$ & \multicolumn{2}{c}{$z_{\rm spec}$}
 & Type & Ref & Comments \\ \hline
VLA0252 & 02:18:06.86 & $-$05:34:15.7 & 15.53$\pm$0.24 & 0.68 (0.67--0.70) &  0.6714 &   & Abs& 1 &  \\
VLA0253 & 02:17:51.63 & $-$05:36:28.3 & 14.71$\pm$0.12 & 0.09 (0.07--0.12) &  0.1034 &   & SB& 1 &  \\
VLA0254 & 02:16:36.81 & $-$05:05:53.7 & 17.55$\pm$0.00 & 0.71 (0.68--0.75) &  0.8052 &   & Strong& 1 &  \\
VLA0255 & 02:17:19.97 & $-$04:33:12.4 &       $>$15.62 & 1.52 (1.43--1.62) &  &   &&  &  \\
VLA0256 & 02:18:28.28 & $-$04:37:36.8 & 19.97$\pm$0.13 & 1.89 (1.46--2.37) & ? &   && 1 &  \\
VLA0257 & 02:18:47.10 & $-$05:13:55.8 & 20.30$\pm$0.06 & 1.66 (1.46--1.92) &  2.0841 &   & BLAGN& 1 &  \\
VLA0258 & 02:16:49.49 & $-$05:18:57.3 &       $>$21.82 & 3.34 (3.16--3.50) &  &   &&  & IRAC detection \\
\hline
\end{tabular}
\end{table*}
\addtocounter{table}{-1}

\begin{table*}
\caption[]{\textit{continued}.}\begin{tabular}{lccccc@{}llcl}
\hline
Name & \multicolumn{2}{c}{IR position (J2000.0)}
& $K$ ($\phi4''$) & $z_{\rm phot}$ & \multicolumn{2}{c}{$z_{\rm spec}$}
 & Type & Ref & Comments \\ \hline
VLA0316 & 02:18:20.08 & $-$05:21:55.3 & 17.46$\pm$0.00 & 0.80 (0.77--0.83) & ? &   && 1 &  \\
VLA0317 & 02:17:46.74 & $-$05:13:14.9 & 20.53$\pm$0.08 & 2.18 (1.82--2.54) &  &   &&  & {\it BVRi\/} affected by nearby star \\
VLA0318 & 02:17:06.34 & $-$05:15:32.7 & 18.32$\pm$0.01 & 1.31 (1.01--1.43) &  1.7415 &   & BLAGN& 6 &  \\
VLA0319 & 02:18:37.68 & $-$05:27:56.6 & 20.43$\pm$0.07 & 1.86 (1.63--2.08) &  &   &&  &  \\
VLA0320 & 02:17:00.04 & $-$04:38:30.9 & 15.44$\pm$0.00 & 0.25 (0.21--0.28) &  0.2939 &   & SB& 1 &  \\
VLA0321 & 02:18:21.78 & $-$04:34:49.9 &       $>$15.51 & 2.00 (1.81--2.17) &  2.2065 &   & NLAGN& 1 &  \\
VLA0322 & 02:17:54.16 & $-$05:27:05.8 & 16.29$\pm$0.00 & 0.70 (0.69--0.72) &  0.6901 &   & Abs& 1 &  \\
\hline
\end{tabular}
\end{table*}
\addtocounter{table}{-1}

\begin{table*}
\caption[]{\textit{continued}.}\begin{tabular}{lccccc@{}llcl}
\hline
Name & \multicolumn{2}{c}{IR position (J2000.0)}
& $K$ ($\phi4''$) & $z_{\rm phot}$ & \multicolumn{2}{c}{$z_{\rm spec}$}
 & Type & Ref & Comments \\ \hline
VLA0380 & 02:16:57.11 & $-$05:31:59.3 & 17.58$\pm$0.00 & 0.68 (0.64--0.72) &  0.4554 &   & BLAGN& 1 &  \\
VLA0381 & 02:18:06.70 & $-$05:25:51.5 & 16.99$\pm$0.00 & 0.68 (0.65--0.71) &  0.6931 &   & Weak& 1 &  \\
VLA0382 & 02:18:45.73 & $-$05:32:55.9 & 16.12$\pm$0.00 & 0.31 (0.26--0.36) &  0.3855 &   & Abs& 3 & merging/projected system \\
VLA0383 & 02:19:49.96 & $-$05:09:06.8 &       $>$15.58 & 1.11 (0.85--1.47) & ? &   && 1 &  \\
VLA0384 & 02:18:40.31 & $-$05:17:32.2 & 21.11$\pm$0.12 & 1.90 (1.54--2.29) &  &   &&  &  \\
VLA0385 & 02:18:51.87 & $-$04:45:18.2 & 18.67$\pm$0.01 & 1.67 (1.59--1.75) &  &   &&  &  \\
VLA0386 & 02:17:25.11 & $-$05:16:17.8 & 20.07$\pm$0.04 & 1.57 (1.37--1.82) & ? &   && 1 &  \\
\hline
\end{tabular}
\end{table*}
\addtocounter{table}{-1}

\begin{table*}
\caption[]{\textit{continued}.}\begin{tabular}{lccccc@{}llcl}
\hline
Name & \multicolumn{2}{c}{IR position (J2000.0)}
& $K$ ($\phi4''$) & $z_{\rm phot}$ & \multicolumn{2}{c}{$z_{\rm spec}$}
 & Type & Ref & Comments \\ \hline
VLA0444 & 02:18:16.78 & $-$05:32:14.3 & 17.35$\pm$0.01 & 0.62 (0.54--0.68) &  0.6935 &   & Strong& 1 &  \\
VLA0445 & 02:17:58.58 & $-$04:49:40.8 & 17.12$\pm$0.00 & 0.50 (0.47--0.53) &  &   &&  &  \\
VLA0446 & 02:17:12.10 & $-$04:52:41.1 & 16.52$\pm$0.00 & 0.29 (0.26--0.32) &  0.3079 &   & SB& 1 &  \\
VLA0447 & 02:16:18.41 & $-$05:06:57.1 & 20.03$\pm$0.04 & 3.31 (3.21--3.51) & ? &   && 1 &  \\
VLA0448 & 02:18:02.79 & $-$05:34:39.1 &       $>$15.79 & 0.78 (0.72--0.83) &  &   &&  &  \\
VLA0449 & 02:18:48.42 & $-$04:38:46.0 & 18.13$\pm$0.02 & 0.71 (0.68--0.76) &  1.1839 &   & BLAGN& 1 &  \\
VLA0450 & 02:16:44.03 & $-$05:24:46.2 & 19.74$\pm$0.03 & 1.67 (1.53--1.81) &  &   &&  & {\it RI\/} optical data \\
\hline
\end{tabular}
\end{table*}

\begin{figure}
\resizebox{\hsize}{!}{\rotatebox{-90}{\includegraphics{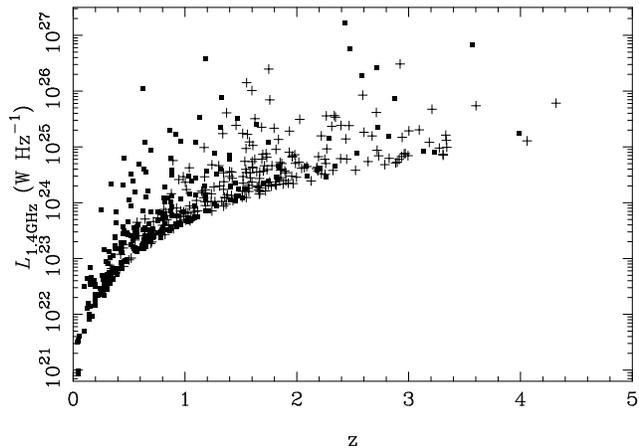}}}
\caption[]{Luminosity--redshift plot for the 100-$\mu$Jy
  sample. Filled squares indicate objects with spectroscopic
  redshifts, while those with photometric redshifts are represented
  by crosses.\label{fig:lzplot}}
\end{figure}

\begin{figure}
\resizebox{\hsize}{!}{\rotatebox{-90}{\includegraphics{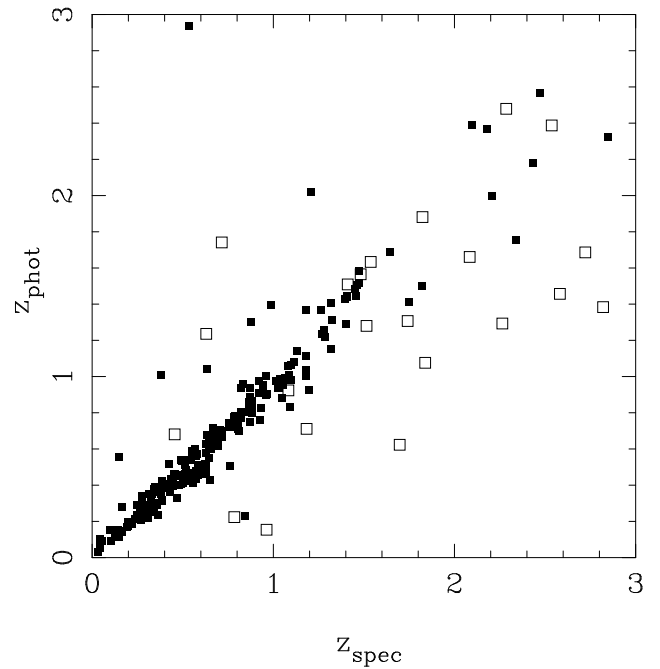}}}
\caption[]{Comparison of spectroscopic and photometric redshifts. Open
  symbols indicate objects with broad emission lines, for which the
  photometric redshift determination is expected to be poor, while
  filled symbols represent all other classes of object.\label{fig:zcompare}}
\end{figure}

\begin{figure}
\resizebox{\hsize}{!}{\rotatebox{-90}{\includegraphics{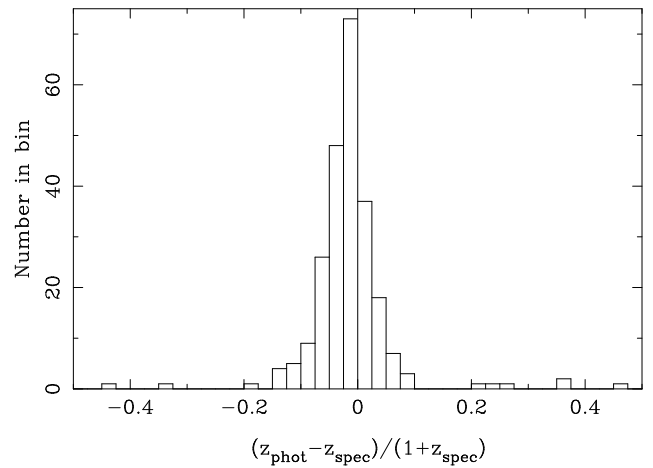}}}
\caption[]{Histogram of normalized photometric redshift errors for the
  objects in Fig.~\ref{fig:zcompare}, excluding the broad-line
  sources. The mean offset is $-$0.023 and the normalized median
  absolute deviation (Brammer et al.\ 2008) is $\sigma_{\rm
    NMAD}=0.035$.\label{fig:zcomparehist}}
\end{figure}

\begin{figure}
\resizebox{\hsize}{!}{\rotatebox{-90}{\includegraphics{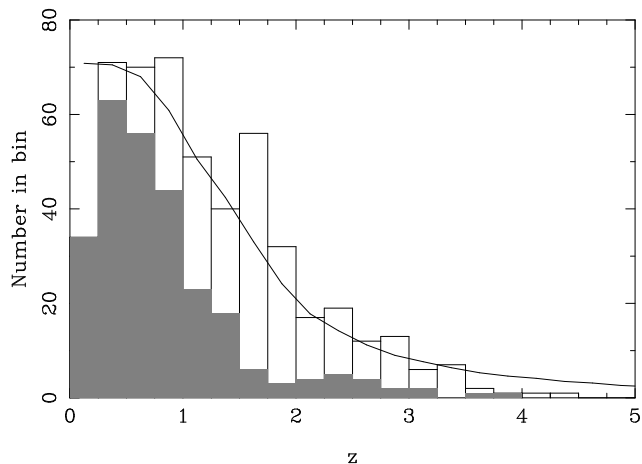}}}
\caption[]{Histogram of redshifts for the 100-$\mu$Jy sample. The
  shaded region of the histogram indicates spectroscopic redshifts
  only, while the open bars represent the photometric redshifts. The
  solid line shows the predicted distribution from the SKA Simulated
  Skies (Wilman et al.\ 2008), renormalized to have the same total
  number of sources.\label{fig:zhistogram}}
\end{figure}

The photometric redshifts were derived using the code \textsc{eazy}
(Brammer, van Dokkum, \& Coppi 2008) after correcting the observed
photometry for the Galactic extinction of $A_V=0.070$ (Schlegel,
Finkbeiner \& Davis 1998) with the Milky Way extinction law of Pei
(1992). All but seven objects have images in eleven filters
(\textit{u$^*$BVRi$'$z$'$JHK\/} plus IRAC channels 1 and 2), while
those seven possess \textit{u$^*$RIJHK\/} and IRAC images. However,
not all these images provide useful photometry for all targets due to
nearby bright sources, and these instances are noted in
Table~\ref{tab:data}.

As noted by Simpson \& Eisenhardt (1999), IRAC photometry can be
important for constraining photometric redshifts through the H$^-$
opacity minimum, which appears as a bump in the spectra of galaxies at
a rest-frame wavelength of 1.6\,$\mu$m. However, this feature can be
absent in AGNs due to thermal emission from hot dust in the obscuring
torus, resulting in a spectral energy distribution that continues to
rise at longer wavelengths and hence an incorrectly high photometric
redshift. After some experimentation it was found that the number of
catastrophic outliers in the $z_{\rm spec}$--$z_{\rm phot}$ plane
could be reduced by adding an additional 10\,per cent uncertainty in
quadrature to the IRAC fluxes. In assigning a photometric redshift to
each object we use the probability-weighted mean redshift, rather than
the most probable redshift. This ensures that the photometric redshift
always lies within the central 68 per cent confidence interval, which
we also list in Table~\ref{tab:data} The luminosity--redshift plane
for our sample in shown in Fig.~\ref{fig:lzplot}.

Fig.~\ref{fig:zcompare} compares the photometric and spectroscopic
redshifts for the 267 sources with spectroscopic redshifts.  Very good
agreement is seen over the range $0<z<1.4$, beyond which the
spectroscopic completeness falls rapidly. The number of catastrophic
outliers also increases at $z_{\rm phot}>1$, but this can be
attributed to a strong bias in the types of sources with spectroscopic
redshifts $z_{\rm spec}>1$. It is difficult to obtain absorption-line
redshifts for distant objects due to the fainter continuum and hence
reduced signal-to-noise ratio, and the redshifting of strong features
(such as the Ca~HK doublet) to wavelengths affected by night sky
emission lines and/or detector fringing. As a result, many of the
sources with spectroscopic redshifts $z>1$ are AGNs whose broad-band
photometry is affected by strong emission lines and dust emission.
Nevertheless, as Fig.~\ref{fig:zcomparehist} shows, the scatter in the
$z_{\rm phot}$--$z_{\rm spec}$ relation is very small, and only 5\,per
cent (11/242) of the sources not classified as `BLAGN' are
catastrophic outliers. Since optically-bright sources and those with
X-ray emission have been preferentially targeted by the spectroscopic
observations described in Section~\ref{sec:otherspec} we expect there
to be few, if any, broad-line AGNs lacking a spectroscopic redshift.

We compare our redshift distribution with that from the SKA Simulated
Skies (S$^3$; Wilman et al.\ 2008), which provides a mock catalogue of
extragalactic radio sources over a $20\times20$\,deg$^2$ area of sky.
The S$^3$ simulations predict 699 sources with
$S_{\rm1.4GHz}>100\,\mu$Jy over our 0.808-deg$^2$ survey area, but we
detected 505 and expect 539 after accounting for incompleteness near
the flux limit (see table~2 of Paper~I). We already noted in Paper~I
that the source counts in the SXDF appear to be significantly lower
than those in some other fields, and we have simply renormalized the
total number of sources in the simulations by 539/699. There is good
agreement between the renormalized model and observations, except in
the lowest redshift bin ($z<0.25$). We speculate that this is due to
radio sources which were resolved in our observations and hence failed
to satisfy the criterion $S_{\rm peak}>100\,\mu$Jy despite having a
total flux density in excess of the threshold. The deficit in this bin
is consistent with the estimated number of sources missed by our
selection on peak flux density (34).

\section{Discussion}

\subsection{The stellar populations of radio sources}
\label{sec:stellarpops}

\begin{figure}
\resizebox{\hsize}{!}{\rotatebox{-90}{\includegraphics{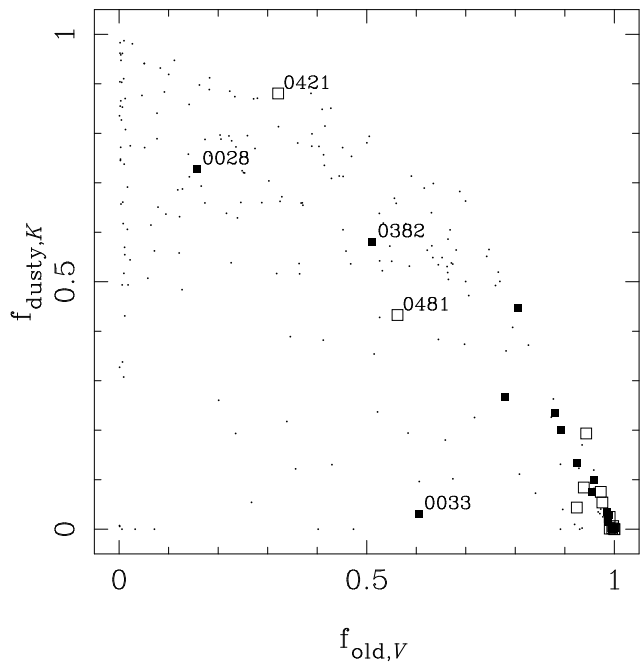}}}
\caption[]{A simple characterisation of the best-fitting stellar
  populations found by \textsc{eazy} for objects with spectroscopic
  redshifts and classifications. The abscissa shows the fraction of
  the rest-frame $V$-band light produced by the three old stellar
  populations, and the ordinate shows the fraction of the rest-frame
  $K$-band light produced by the young dust-reddened population. Large
  filled symbols represent objects without emission lines (`Abs' in
  Table~\ref{tab:data}), and open symbols are objects with weak line
  emission (`Weak' in Table~\ref{tab:data}). Small points are sources
  with other classifications. Objects discussed in the text are
  labelled with their source numbers.\label{fig:oldyoung}}
\end{figure}

\textsc{eazy} derives its photometric redshifts by fitting a linear
combination of six synthetic stellar population spectra to the
observed broadband fluxes. These spectra have non-trivial star
formation histories but three can be considered `old' and three
`young', including one dusty starburst spectrum. We choose to
characterise the stellar population of the best-fitting model with two
numbers: the fraction of rest-frame $V$-band light which arises from
old stars (i.e., the fraction from the three `old' templates),
$f_{{\rm old},V}$, and the fraction of rest-frame $K$-band light
produced by the dusty starburst template, $f_{{\rm dusty},K}$. When we
compare this to the spectroscopic classifications
(Fig.~\ref{fig:oldyoung}), there is a clear correspondence between
objects classified as `old' and those lacking emission lines. Outliers
are labelled in the figure, and their locations are readily
understood. The photometric measurements of VLA~0033 and VLA~0382 are
contaminated by nearby objects (a $z>1$ gravitationally-lensed arc in
the case of VLA~0033; see Geach et al.\ 2007 for more
details). VLA~0028 is unusual in that, while clearly resolved at
longer wavelengths, it appears stellar in the $u^*$ and $B$
filters. This can also be seen in the unusual `concave' nature of its
spectrum in Fig.~\ref{fig:spectra}. It may be a chance superposition
of an early-type Galactic star with the radio galaxy, or the galaxy
may harbour a BL~Lac nucleus.

\begin{figure}
\resizebox{\hsize}{!}{\rotatebox{-90}{\includegraphics{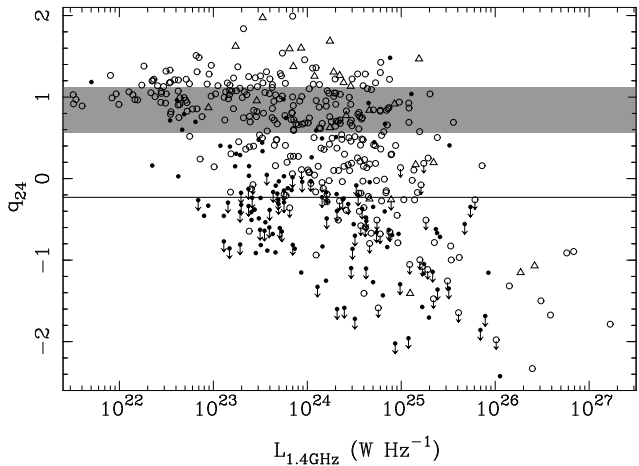}}}
\caption[]{Plot of $k$-corrected $q_{24}$ values against 1.4-GHz
  luminosity density. Objects classified as BLAGN are shows as
  triangles. Filled symbols indicate galaxies whose stellar
  populations are classified as `old' according to the criterion of
  Section~\ref{sec:stellarpops}. The shaded region shows the
  $\pm1\sigma$ range for star-forming galaxies from Appleton et
  al.\ (2004), while the horizontal line indicates Ibar et al.'s
  (2008) boundary between radio-loud and radio-quiet sources, which we
  also adopt.\label{fig:q24old}}
\end{figure}

The criterion $f_{{\rm old},V}>0.82$ selects 90\% (52/58) of sources
classified as `Weak' or `Abs', but only 11\% (21/192) of other
sources, and we therefore use this criterion to identify host galaxies
without any significant recent star-formation activity or strong
unobscured accretion, which we term `old'. We independently assess
this criterion by using 24-$\mu$m imaging from the SpUDS
\textit{Spitzer\/} survey (PI: J.~Dunlop) to identify objects with
strong star-formation or accretion activity. Due to the high source
density of this image, we follow the method of Rodighiero et
al.\ (2006) by running the \textsc{clean} algorithm (H\"{o}gbom 1974)
on the image to reduce the effects of confusion from the Airy rings.
The \textsc{clean}ed image is then convolved with a Gaussian of FWHM
5.5$''$ and added to the residual image. A 24-$\mu$m flux density is
measured for each radio source by fitting a Gaussian at its location
in this image. Where this fit was affected by a nearby 24-$\mu$m
source we employed a `search and destroy' method to eliminate such
sources before fitting a point spread function to the counterpart. We
$k$-correct the 24-$\mu$m photometry using Silva et al.'s (1998) model
fit to the spectral energy distribution of M82, which we convolve with
the transmission profile of the MIPS filter. We have also undertaken
all subsequent analysis assuming an AGN-like spectral energy
distribution that is flat in $\nu L_\nu$ and find no significant
differences in our results. After $k$-correcting the 1.4-GHz radio
flux, we compute the ratio of mid-infrared to radio flux $q_{24} =
\log (S_{\rm24\mu m}/S_{\rm1.4GHz})$. This quantity identifies
radio-loud AGNs which lie away from the tight infrared--radio
correlation that is obeyed by both star-forming galaxies and
radio-quiet AGNs (e.g., Sopp \& Alexander 1991).

In Fig.~\ref{fig:q24old} we plot $q_{24}$ against radio luminosity and
indicate which galaxies have been classified as `old' by our
criterion\footnote{Thirteen objects from the sample are not covered by
  the MIPS imaging and have therefore been excluded from this analysis
  without biasing the results.}. We also plot a line at
$q_{24}=-0.23$, which Ibar et al.\ (2008) used to select radio-loud
galaxies: this is equivalent to a radio-loudness parameter
$R\approx30$ (cf.\ Kellermann et al.\ 1989) if we assume the
relationship between 24-$\mu$m and $B$-band luminosities implied by
the radio-loud QSO spectral energy distribution of Shang et
al.\ (2011). It is clear that old galaxies do not represent a random
sample of the population, with the fraction of old galaxies among the
radio-loud population decreases from nearly 100\,per cent at
$L_{\rm1.4GHz}\sim10^{23}\rm\,W\,Hz^{-1}$ to almost zero at
$L_{\rm1.4GHz}>10^{26}\rm\,W\,Hz^{-1}$. This can be understood as a
result of the Hine \& Longair Class~A fraction increasing with
luminosity, since these objects possess a strong non-stellar ionizing
continuum (which may be obscured) and also frequently show evidence
for recent star formation (see Tadhunter et al.\ 2011 and references
therein), both of which would prevent the host galaxy from being
classified as old. Class~B sources dominate at lower radio
luminosities and these sources usually display the spectra of old
stellar populations.

\begin{figure}
\resizebox{\hsize}{!}{\rotatebox{-90}{\includegraphics{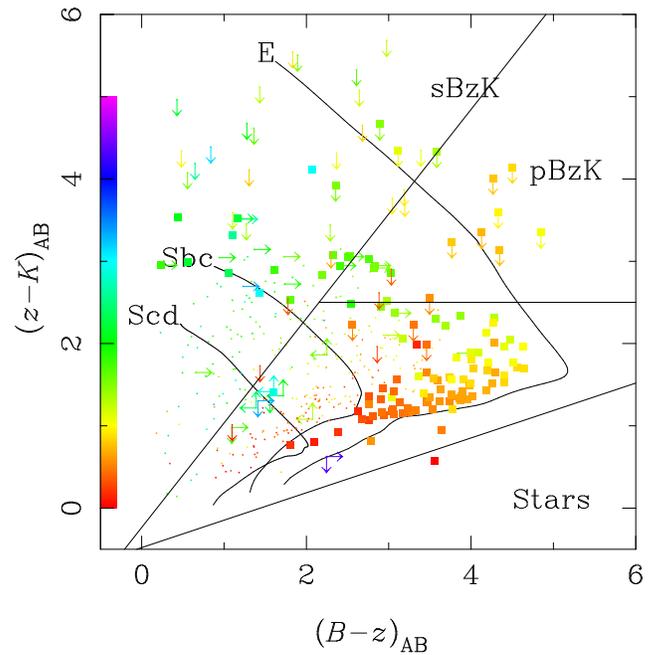}}}
\caption[]{Colour--colour plot for all radio sources, colour-coded
  according to their redshifts (spectroscopic where available,
  otherwise photometric), as indicated by the bar on the
  $y$-axis. Large filled symbols indicate objects classified as `old'
  according to the criterion in the text. The parameter space is split
  in the manner of Daddi et al.\ (2004), after applying photometric
  corrections to account for filter differences (W.~Hartley, private
  communication). The loci of non-evolving galaxy templates from
  Coleman, Wu, \& Weedman (1980) are plotted (W.~Hartley, private
  communication). \label{fig:bzk}}
\end{figure}

We compare our classification with the locations of sources in the
\textit{BzK\/} colour--colour diagram of Daddi et al.\ (2004) in
Fig.~\ref{fig:bzk}, where we also plot the loci of the non-evolving E,
Sbc, and Scd galaxy templates from Coleman, Wu, \& Weedman
(1980). This diagram is used to select galaxies at $z\ga1.4$ and
classify them as either passively-evolving (pBzK) or star-forming
(sBzK) objects.  Surprisingly, the pBzK region of this diagram
contains very few sources, with the locus of `passive' radio galaxies
lying between the E and Sbc curves. Since passive galaxies at $z>1.4$
have very red $(B-z)$ colours, even a small amount of ultraviolet
luminosity will cause galaxies to migrate from the pBzK region into
the sBzK region of the diagram, as is apparent from the presence of
`old' galaxies in the sBzK region of the diagram. It therefore appears
that there is some ultraviolet emission associated with AGN activity
even in sources which display no line emission but we cannot say
whether this arises from stellar or non-stellar processes. We conclude
from Figures~\ref{fig:q24old} and \ref{fig:bzk} that our criterion for
selecting old galaxies is reliable although it is not as severe as the
pBzK criterion.

\subsection{The K--z relation for faint radio sources}
\label{sec:kz}

It is well-known that radio galaxies (i.e., the host galaxies of
radio-loud AGNs) follow a tight locus in the $K$-band Hubble diagram,
the so-called $K$--$z$ relation (Lilly \& Longair 1984; Jarvis et
al.\ 2001; De Breuck et al.\ 2002; Willott et
al.\ 2003). Historically, the $K$-band magnitudes used in this diagram
have been measured in a fixed metric aperture of 63.9\,kpc diameter,
which corresponds to $\sim8''$ at high redshift. However, the compact
nature of distant radio galaxies means that the magnitudes are almost
universally measured in smaller apertures to reduce the photometric
uncertainty, and then an offset is applied based on an assumed curve
of growth (the prescription of Eales et al.\ 1997 is frequently
followed). While this does, in principle, enable objects from
different samples to be plotted on the same plot, it is only valid if
all objects obey the same curve of growth, or have all had their
photometry measured in identical angular apertures so that the same
offset is applied to all. The Eales et al.\ (1997) prescription was
derived for bright radio sources which are known to be massive
ellipticals, but this is clearly not applicable to sources from
fainter radio samples such as the one studied here. We therefore
follow Bryant et al.\ (2009a) in using apertures of the same angular
size (4-arcsecond diameter) at all redshifts. Once again, we apply a
correction to the 2MASS photometry and, following the method in
Section~\ref{sec:IRphotom}, we determine $K_{\rm2MASS}(\phi4'')-K_{\rm
  UDS}(\phi4'')=0.35$. These measurements are also quoted in
Table~\ref{tab:data}.

\begin{figure}
\resizebox{\hsize}{!}{\rotatebox{-90}{\includegraphics{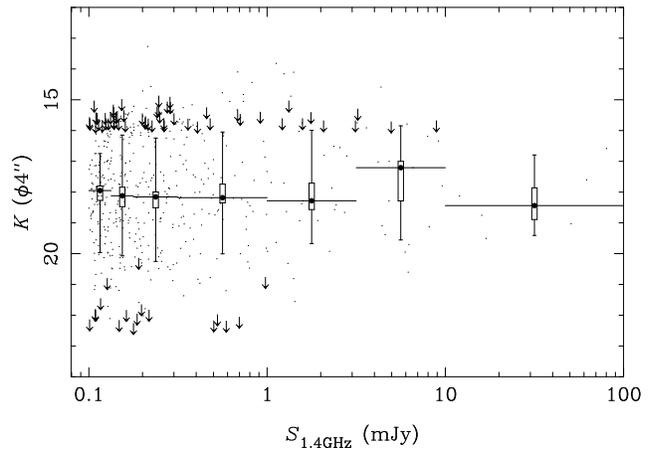}}}
\caption[]{Plot of $K$-band magnitude ($\phi4''$) against 1.4-GHz
  radio flux density. Large symbols show the median $K$ magnitude in
  each flux density bin, derived from the unbiased subset of objects
  with UDS photometry. The vertical error bars show the 16th and
  84th percentiles of the distribution in each bin, while the
  boxes show the 1$\sigma$ uncertainty on the median estimated
  from 500 bootstrapped samples.\label{fig:fluxplot}}
\end{figure}

In Paper~I we showed that there was no significant change in the
optical colours of radio sources over the radio flux density range
$0.1{\rm\,mJy} \leq S_{\rm1.4GHz} < 100\rm\,mJy$, and inferred that,
although present, a population of star-forming galaxies could not be
dominating at the faintest fluxes. Fig.~\ref{fig:fluxplot} shows that
there is also no significant variation in the median $K$-band
magnitudes of radio sources as a function of radio flux density.

\begin{figure}
\resizebox{\hsize}{!}{\rotatebox{-90}{\includegraphics{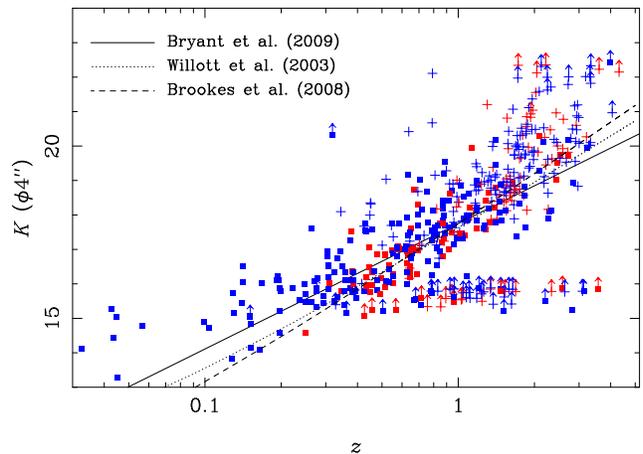}}}
\caption[]{$K$-band Hubble diagram for the SXDF 100-$\mu$Jy
  sample. Squares indicate spectroscopic redshifts, while crosses show
  objects with photometric redshift estimates only. Red and blue
  points indicate radio-loud and radio-quiet sources,
  respectively. The lines indicate the $K$--$z$ relations from Bryant
  et al.\ (2009b; fit over the entire redshift range), Willott et
  al.\ (2003), and Brookes et al.\ (2008). The relations from the
  latter two papers have been corrected to our 4-arcsec aperture using
  the prescription of Eales et al.\ (1997).\label{fig:kz}}
\end{figure}

The original $K$--$z$ relation was determined for the most powerful
radio sources from the 3C catalogue (Lilly \& Longair 1984) and, as
deeper radio surveys have been undertaken, a number of papers have
appeared in the literature investigating the variation in the $K$--$z$
relation with radio luminosity. Eales et al.\ (1997) and Willott et
al.\ (2003) showed that galaxies from the fainter 6C and 7C catalogues
were 0.6\,magnitudes fainter at a given redshift which, after
excluding a direct correlation with AGN luminosity (Leyshon \& Eales
1998; Simpson, Rawlings \& Lacy 1999), was determined to be due to a
difference in the stellar luminosities of their host
galaxies. However, the faintest of these samples (7C) has a radio flux
limit approximately 1000 times brighter than that of our sample,
compared to a factor of only 20 between 7C and 3C. Vardoulaki et
al.\ (2012) extended this analysis a further factor of $\sim30$ using
the bright SXDF sample of Paper~II and identified sources
significantly fainter than the $K$--$z$ relation. A problem with the
analysis of faint samples is that the strong correlation between
emission-line and radio luminosity (e.g., Rawlings \& Saunders 1991)
means that sources from fainter samples will have lower line fluxes,
frequently resulting in the (often marginal) detection of only a
single line.  While in some cases this line can be confidently
identified due to the absence of other lines, many authors have used
the observed $K$ magnitude to support an identification (we stress
that Vardoulaki et al.\ do not do this). This leads to circular
reasoning because objects are placed on the $K$--$z$ at least partly
under the assumption that the $K$--$z$ relation is reliable, and these
objects are then used to support that very assumption. We reiterate
that all the spectroscopic redshifts presented here are based solely
on the results of the spectroscopy, and in Fig.~\ref{fig:kz} we plot
the $K$-band Hubble diagram for our sample. We include on this plot
the $K$--$z$ relation of Bryant et al.\ (2009), plus those presented
by Willott et al.\ (2003) and Brookes et al.\ (2008), both of which
have been corrected to our fixed angular-diameter aperture using the
same method from Eales et al.\ (1997) that they used to correct their
data to the standard 63.9-kpc aperture. The relation of Brookes et
al.\ differs from the other two in that the least-squares fit
minimized the residuals in $z$, rather than $K$, since $K$ is the
independent variable when the relation is used to infer redshifts. It
is also derived from a radio source sample that is fainter than those
of Willott et al.\ and Bryant et al., although the radio flux limit is
still 72 times brighter than ours. Fig.~\ref{fig:kz} shows quite
clearly that all three $K$--$z$ relations fail to provide a good fit
to our data at $z\ga1$, where the sources are systematically fainter
than the relation.

\begin{figure}
\resizebox{\hsize}{!}{\rotatebox{-90}{\includegraphics{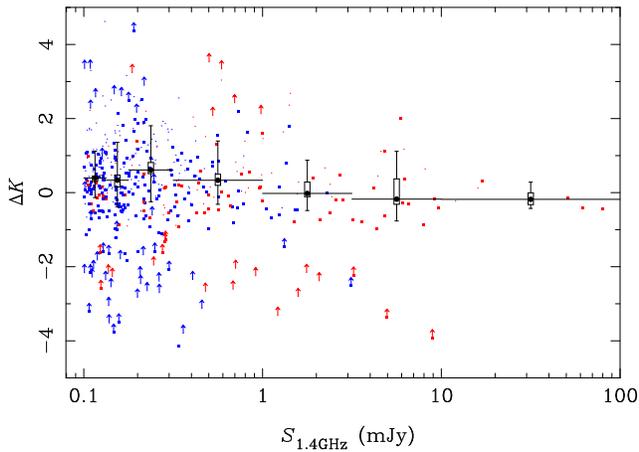}}}
\caption[]{Plot of the deviation from the $K$--$z$ relation of
  Bryant et al.\ (2009b) against 1.4-GHz radio flux density. Filled squares
  show objects with spectroscopic redshifts. The large filled circles
  show the median $K$ magnitude in each flux density bin, derived from
  the unbiased subset of objects with UDS photometry. The vertical
  error bars and boxes have the same meaning as in
  Fig.~\ref{fig:fluxplot}.\label{fig:deltak}}
\end{figure}

\begin{table}
\caption[]{Median $K$ magnitude offset from the $K$--$z$ relations of
  Fig.~\ref{fig:kz} for our $z>1$ sources. The error bars indicate the
  1$\sigma$ uncertainty in the median from 500 bootstrapped
  samples.\label{tab:deltak}}
\begin{center}
\begin{tabular}{r@{--}rrrr}
\hline
\multicolumn{2}{c}{$S_{\rm1.4GHz}$} & \multicolumn{3}{c}{Median offset from
  $K$--$z$ relation} \\
\multicolumn{2}{c}{(mJy)} & \multicolumn{1}{c}{Bryant} &
\multicolumn{1}{c}{Willott} & \multicolumn{1}{c}{Brookes} \\ \hline
10.000 & 100.000 & $-0.13^{+0.25}_{-0.10}$ & $-0.23^{+0.29}_{-0.07}$ &
$-0.42^{+0.26}_{-0.08}$ \\
3.162 & 10.000 & $0.64^{+0.47}_{-0.19}$ & $ 0.61^{+0.33}_{-0.23}$ &
$0.41^{+0.20}_{-0.25}$ \\
1.000 & 3.162 & $0.28^{+0.19}_{-0.31}$ & $ 0.35^{+0.02}_{-0.29}$ &
$0.08^{+0.08}_{-0.06}$ \\
0.316 & 1.000 & $0.52^{+0.16}_{-0.17}$ & $ 0.40^{+0.22}_{-0.11}$ &
$0.18^{+0.17}_{-0.12}$ \\
0.178 & 0.316 & $0.96^{+0.24}_{-0.11}$ & $ 0.92^{+0.21}_{-0.13}$ &
$0.73^{+0.16}_{-0.16}$ \\
0.133 & 0.178 & $0.69^{+0.23}_{-0.22}$ & $ 0.69^{+0.13}_{-0.24}$ &
$0.44^{+0.16}_{-0.27}$ \\
0.100 & 0.133 & $0.57^{+0.05}_{-0.12}$ & $ 0.52^{+0.06}_{-0.14}$ &
$0.25^{+0.08}_{-0.10}$ \\
\hline
\end{tabular}
\end{center}
\end{table}

We also show the classification of sources as radio-loud or
radio-quiet in this figure, based on the analysis of
Section~\ref{sec:stellarpops}. For sources outside the SpUDS region or
with $q_{24}$ upper limits larger than $-$0.23, where we cannot make a
conclusive determination as to whether they are radio-loud or
radio-quiet, we use the result of Fig.~\ref{fig:q24old} to identify
`old' galaxies as being radio-loud.  Although the redshift
distributions of the radio-loud and radio-quiet populations differ,
with the lowest-redshift sources being exclusively radio-quiet, it is
clear that this deviation from the $K$--$z$ relation is not simply
caused by the presence of radio-quiet sources in the sample.

We investigate further by plotting the deviation from the $K$--$z$
relation of Bryant et al.\ (2009b) against radio flux density in
Fig.~\ref{fig:deltak}. Using only the unbiased set of objects within
the UDS region and including magnitude limits, the Kendall's rank
correlation statistic (Brown, Hollander \& Korwar 1974; Isobe,
Feigelson \& Nelson 1986) indicates the presence of an anticorrelation
between these quantities at greater than 95\,per cent significance. If
only those objects with secure spectroscopic redshifts are used, the
level of significance is slightly lower (92\,per cent), but this
subsample suffers from numerous biases that are hard to quantify, such
as a bias against objects with low emission-line fluxes, or those in
the `redshift desert'.

This result is not caused by our choice of $K$--$z$
relation. Table~\ref{tab:deltak} shows the median offset of our $z>1$
sources from each of the three relations. We consider only the
highest-redshift sources since these are the ones for which this form
of redshift estimator is most likely to be used, and the differences
between the different relations is dominated by aperture corrections
at low redshift. We also undertake this analysis for radio-loud and
radio-quiet sources separately, and find no significant difference
between the two samples. We find that the sub-mJy radio sources are
systematically fainter than the $K$--$z$ relations by up to a
magnitude, resulting in the overestimation of redshifts and hence
luminosities (by up to a factor of two). In reality, however, the use
of a combination of spectroscopic redshifts and those estimated from
the $K$--$z$ relation will produce larger errors. As
Fig.~\ref{fig:deltak} clearly shows, the distribution of deviations is
not symmetric about the median but has a tail to even fainter
magnitudes, and it is these objects which are less likely to have
measured spectroscopic redshifts. Results that stem from the use of
`single-band photometric redshifts', or from spectroscopic redshifts
derived from a single emission line plus a broad-band magnitude, must
therefore be treated with extreme caution. This bias can only be
eliminated by estimating the redshift of a source based on its $K$
magnitude and radio flux density, and also whether or not spectroscopy
has been attempted, using an extension of the method of Cruz et
al.\ (2007). However, this requires a faint radio sample with 100\,per
cent reliable redshifts and that does not yet exist.

\subsection{Evolution of the radio luminosity function}
\label{sec:rlf}

\begin{figure*}
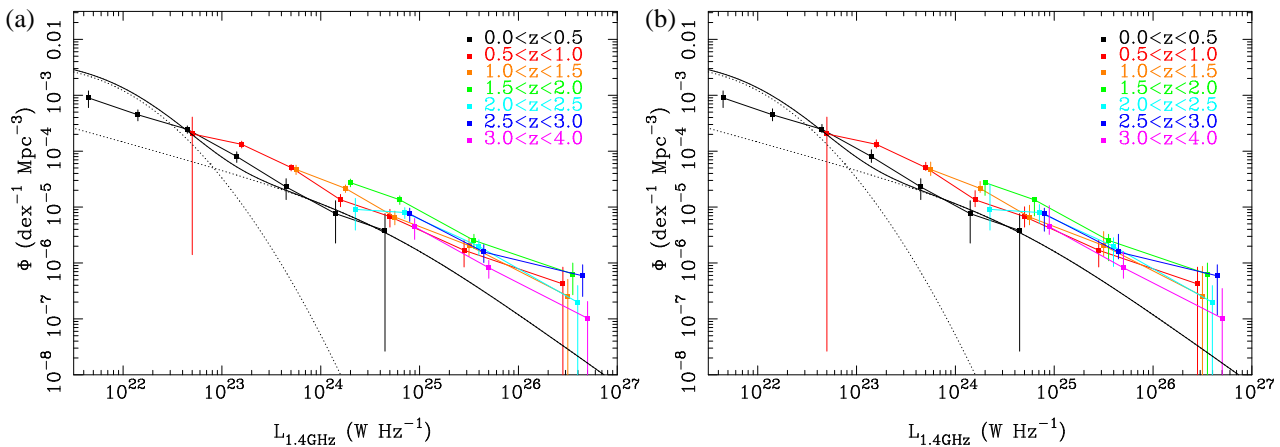

\begin{picture}(0,0)
\put(0,-6){\large(a)}
\put(242,-6){\large(b)}
\end{picture}
\resizebox{\colwidth}{!}{\rotatebox{-90}{\includegraphics{plot_lf.ps}}}
\resizebox{\colwidth}{!}{\rotatebox{-90}{\includegraphics{plot_lf2.ps}}}
\caption[]{Evolution of the radio luminosity function with redshift
  for the SXDF sample. The left-hand plot ignores photometric redshift
  uncertainties, while the right-hand plot incorporates these using
  1000 Monte Carlo realizations as described in the text, with the
  error bars indicating the range between the 16th and 84th percentile
  for each data point. Different colour symbols denote different
  redshift ranges as indicated in the key. Bins are of of width
  $\Delta\log L_{\rm1.4GHz}=0.5$ for
  $L_{\rm1.4GHz}\leq10^{25}$\,W\,Hz$^{-1}$ and $\Delta\log
  L_{\rm1.4GHz}=1.0$ at higher luminosities. Points have only been
  plotted where there is at least one object in the bin, and points
  within the same luminosity bin representing different redshift bins
  are offset horizontally for clarity. The black solid line is the
  parametrized fit to the local radio luminosity function of Mauch \&
  Sadler (2007), and the dotted curves show the separate contributions
  from star-forming galaxies and AGNs.\label{fig:lfplot}}
\end{figure*}

At the flux densities we are studying, the extragalactic radio source
counts are comprised of (at least) four populations: star-forming
galaxies, radio-quiet AGNs, and two types of radio-loud AGN. Due to
their physically distinct natures, these populations need not undergo
similar evolution and therefore an accurate parametrization of the RLF
and its evolution is not a realistic proposition. Instead we construct
binned luminosity functions using the method of Page \& Carrera (2000)
and present these in Fig.~\ref{fig:lfplot}. The error bars in
Fig.~\ref{fig:lfplot}(a) account for only the Poisson errors and
ignore the photometric redshift uncertainties, while we have
propagated these uncertainties to produce Fig.~\ref{fig:lfplot}(b) by
running 1000 Monte Carlo realisations where each source without a
spectroscopic redshift is assigned a photometric redshift based on the
probability density function returned by {\sc eazy}. We calculate the
asymmetric 1-$\sigma$ error bar from the 16th and 84th percentiles of
the distribution in each bin and add this in quadrature to the Poisson
uncertainty. Unsurprisingly, the contribution to the error budget from
the photometric redshift uncertainty increases with redshift as the
hosts become fainter and the photometric errors grow. There will also
be a further contribution to the error budget from our lack of
knowledge of the sources' spectral indices and a systematic effect
caused by spectral curvature (e.g., Jarvis \& Rawlings 2000), although
we presently lack sufficiently deep multifrequency radio data to
quantify these effects.  Accounting for the photometric redshift
uncertainties is important since we can construct the RLF for all
radio sources (except the seven sources described in
Section~\ref{sec:exclusion} whose photometry is contaminated by nearby
objects), even if they lack a formal optical or infrared
counterpart. We have not corrected for incompleteness towards
marginally-resolved sources near the radio flux limit since
Fig.~\ref{fig:zhistogram} suggests that this only affects objects in
the lowest redshift bin. Indeed, the difference between our $z<0.5$
RLF and that of Mauch \& Sadler (2007) in Fig.~\ref{fig:lfplot}
supports this suggestion.

The radio source population shows strong evolution out to $z\sim1$
over the entire luminosity range probed by our sample, with the space
density of these objects increasing by a factor of 3 over this
lookback time. The evolution is stronger for higher luminosity radio
sources when our results are compared with the local RLF. This is
consistent with the evolution found by Sadler et al.\ (2007) in their
comparison between the 6dFGS and 2SLAQ radio galaxy samples. At higher
redshift, there appears to be a slowing down of this evolution and
there is a hint of negative evolution at higher redshift, with the
space densities at $z>2$ being systematically lower than at later
cosmic times, albeit at very low significance. This is not
straightforward to interpret due the composition of the radio
population evolving at these intermediate luminosities. Locally,
sources with $L_{\rm1.4GHz}\sim10^{24{\rm-}26}\rm\,W\,Hz^{-1}$ belong
to Hine \& Longair's Class~B and are massive ellipticals engaged in a
constant cycle of heating infalling gas so as to prevent additional
star formation, operating in what Croton et al.\ (2006) dub
`radio-mode' feedback. However, as lookback time increases, the strong
evolution in the number of rapidly star-forming galaxies and luminous
radio-quiet AGNs causes these to become significant contributors. For
example, $L_{\rm1.4GHz}=10^{24}\rm\,W\,Hz^{-1}$ corresponds to a
star-formation rate of $\sim200\rm\,M_\odot\,yr^{-1}$ in
$M>5\rm\,M_\odot$ stars (Condon 1992), or $\sim
540\rm\,M_\odot\,yr^{-1}$ in all stars above 0.08\,M$_\odot$ using the
Kroupa (2001) initial mass function and, while such objects are rare
in the local Universe, they are found in submillimetre surveys with a
redshift distribution that peaks at $z\sim2$ (e.g., Chapman et
al.\ 2005).

\begin{figure*}
\begin{picture}(0,0)
\put(0,-6){\large(a)}
\put(242,-6){\large(b)}
\end{picture}
\resizebox{\colwidth}{!}{\rotatebox{-90}{\includegraphics{plot_lf2_rl.ps}}}
\resizebox{\colwidth}{!}{\rotatebox{-90}{\includegraphics{plot_lf2_rq.ps}}}
\caption[]{Evolution of the radio luminosity function for (a)
  radio-loud and (b) radio-quiet galaxies. We define galaxies as
  radio-loud if they have $q_{24}<-0.23$ or have $q_{24}$ upper limits
  greater than this value and have been classified as `old' in
  Fig.~\ref{fig:oldyoung}.\label{fig:lfplot_q24}}
\end{figure*}

\begin{figure}
\resizebox{\colwidth}{!}{\rotatebox{-90}{\includegraphics{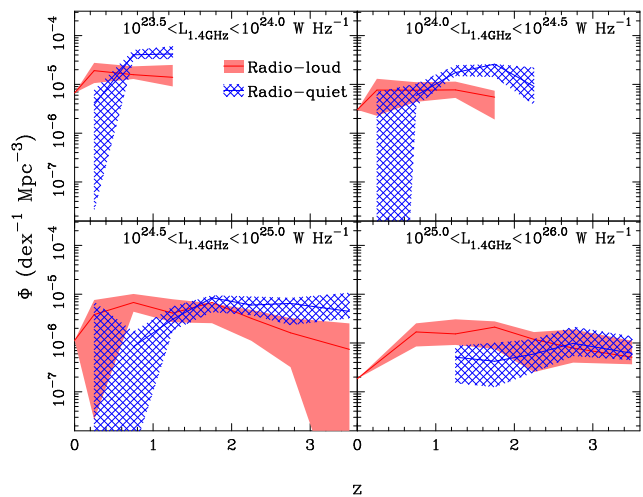}}}
\caption[]{Plots of the redshift evolution of the space density of
  radio sources from Fig.~\ref{fig:lfplot_q24}, binned by
  luminosity. The shaded regions show the 1-$\sigma$ uncertainties in
  each bin. The radio-loud curves have been extended to $z=0$ using
  the Mauch \& Sadler (2007) AGN radio luminosity function, since this
  is dominated locally by radio-loud objects at these
  luminosities.\label{fig:zlfplot}}
\end{figure}

We next investigate the evolution of the radio-loud and radio-quiet
subpopulations independently, again running 1000 Monte Carlo trials to
quantify the effect on the RLF of the photometric redshift
uncertainties and reclassifying objects as radio-loud or radio-quiet
based on their new $k$-corrected $q_{24}$ values. This results in the
luminosity functions shown in Fig.~\ref{fig:lfplot_q24}. We see first
that this split agrees fairly well at low redshift with the separate
AGN and star-formation RLFs of Mauch \& Sadler (2007), despite the
fact that radio-quiet AGNs are grouped with star-forming galaxies in
our classification scheme but with radio-loud AGNs by Mauch \&
Sadler. Such objects comprise only a few per cent of the 6dFGS--NVSS
sample and so the RLF is not particularly sensitive to their
classification except at the bright end of the radio-quiet RLF, where
AGNs dominate over star-forming galaxies (we note also the recent
claim that the radio emission from some radio-quiet AGNs may be
powered by star formation; Kimball et al.\ 2011). It is also clear
that the evolution at intermediate radio luminosities is driven by the
radio-quiet sources, as we find zero or negative evolution among the
radio-loud sources with $L_{\rm1.4GHz}\la10^{24}\rm\,W\,Hz^{-1}$. This
agrees with the finding of Padovani et al.\ (2011) and
Fig.~\ref{fig:zlfplot} shows very clearly that the space density of
radio-quiet sources can exceed that of radio-loud ones, even at radio
luminosities close to the Fanaroff--Riley break. At
$L_{\rm1.4GHz}\approx10^{24}\rm\,W\,Hz^{-1}$ the radio-quiet sources
outnumber the radio-loud ones by $z\sim1$ as one looks to earlier
cosmic epochs, whereas this does not happen until $z\sim2$ for sources
with $L_{\rm1.4GHz}\approx10^{25}\rm\,W\,Hz^{-1}$.

Fig.~\ref{fig:zlfplot} also shows hints of a decline in the space
density of radio-loud galaxies at $z\ga1$, and this decline appears to
happen later (i.e., at lower redshift) for lower-luminosity
objects. Similar claims have been made by Waddington et al.\ (2001)
and Rigby et al.\ (2011) and our results are quantitatively consistent
with these claims, although we fail to robustly identify the turnover
at $z\sim1$ for sources with $L_{\rm1.4GHz}>10^{25}\rm\,W\,Hz^{-1}$
claimed by Rigby et al. Any decline of the radio-quiet population is
less severe and/or happens at higher redshift where we lack
sensitivity. Although we do not separate our radio-quiet sources into
AGN and star-forming subsamples, we note that Padovani et al.\ (2011)
found very similar evolution for these two samples in their study of
radio sources in the \textit{Chandra\/} Deep Field South, and we find
similarly strong evolution. At luminosities where star-forming
galaxies dominate locally ($L_{\rm1.4GHz}\sim10^{23}\rm\,W\,Hz^{-1}$)
we see evolution of the space density that is consistent with the
increase in the cosmic star formation density (e.g., fig.~1 of Hopkins
\& Beacom 2006). At higher luminosities, where the population is
composed of radio-quiet AGNs, the evolution over the range $0.5\la
z\la2$ is consistent with the rapid increase in the number of
optically-selected QSOs (e.g., Croom et al.\ 2004). Adopting the
radio-quiet QSO spectral energy distribution of Shang et al.\ (2011),
radio-quiet QSOs with $L_{1.4GHz}=10^{25}\rm\,W\,Hz^{-1}$ should have
$M_{b_J}\approx-27$ but the space density of the former is several
times greater than that of the latter. This is readily explained by a
combination of two factors. First, the radio-selected sample includes
not just optically-bright QSOs but also obscured AGNs. These can be
`Type 2' objects of the unified model (e.g., VLA0074) or
`host-obscured' objects that Mart\'{\i}nez-Sansigre et al.\ (2008)
claim dominate the accretion onto supermassive black holes. Second,
radio selection will inevitably find the `radio-intermediate' objects
that are believed to be Doppler-boosted radio-quiet sources (Falcke,
Sherwood \& Patnaik 1996) and therefore have fainter optical
luminosities.

The luminosity-dependent turnover in the space density of radio-loud
AGNs appears to be similar to the finding of Fiore et al.\ (2003) that
the space density of luminous X-ray-selected AGNs peaks at a higher
redshift than the space density of lower-luminosity sources. Both
results have been explained in the context of `downsizing' but radio
selection produces a more complex sample than shorter-wavelength
selection methods. This is because other selection methods (e.g.,
optical or X-ray) can only find supermassive black holes that are
undergoing radiatively-efficient accretion (`quasar mode') and there
is a correlation between luminosity and black hole mass because the
range of black hole masses is greater than the spread in Eddington
ratios (e.g., Kelly et al.\ 2010). However, radio selection also finds
black holes that are accreting in a radiatively-inefficient mode
(`radio mode') so this correlation is broken as one moves below the
Hine \& Longair boundary.

To understand this result we first turn to the work of Best et
al.\ (2005a,b) who determined the fraction, $f_{\rm radio-loud}$, of
galaxies in the local Universe brighter than some 1.4-GHz luminosity,
$L$, as a function of black hole mass, estimated from the stellar
velocity dispersion. They found that more massive black holes (in more
massive galaxies) were more likely to host the most luminous radio
sources and derived an expression for $f_{\rm radio-loud}$ as a
function of $L$ and black hole mass, $M$.  From this definition, it
follows that the probability that a galaxy with a given black hole
mass has a radio luminosity between $L$ and $L+dL$ is given by
\begin{equation}
p(L|M) \, dL = - \frac{\partial f_{\rm radio-loud}}{\partial L} \, dL
\end{equation}
and hence the overall RLF is given by
\begin{equation}
\Phi(L) = \int \phi(M) \, p(L|M) \, dM \, ,
\end{equation}
where $\phi(M)$ is the black hole mass function. However, Best et
al.\ (2005b) were able to write the functional form of $f_{\rm
  radio-loud}$ in a manner that is separable in $L$ and $M$ and, for
any such separable form,
\begin{equation}
f_{\rm radio-loud} = f(L) \, g(M)
\end{equation}
we can write
\begin{equation}
\Phi(L) = - f'(L) \int \phi(M) \, g(M) \, dM \, ,
\label{eqn:best}
\end{equation}
where $f'(L)$ is the first derivative of $f(L)$ with respect to
luminosity. The shape of the RLF is therefore independent of the black
hole mass distribution if Best et al.'s expression for $f_{\rm
  radio-loud}$ is true at all redshifts. However, their sample was
dominated by Hine \& Longair Class~B objects and the expression might
not be applicable to Class~A objects due to the different mode of
accretion. Nonetheless, in a general sense, if the likelihood of a
black hole powering a radio-loud AGN increases with its mass and
$\phi(M)$ has a Schechter-like form with a sharp cutoff at high mass
(e.g., Graham et al.\ 2007), it is black holes with masses near the
characteristic mass, $M^*$, that drive the overall shape of the RLF.

In the most recent theoretical models for the evolution of
supermassive black holes, radio jets are powered by the
Blandford--Znajek mechanism (Blandford \& Znajek 1977) and the radio
jet power is a function of the black hole spin (e.g., Fanidakis et
al.\ 2011; Mart\'{\i}nez-Sansigre \& Rawlings 2011). The panels in
Fig.~\ref{fig:zlfplot} therefore represent different black hole spins,
with the lowest luminosity objects having the lowest spins. The
luminosity-dependent turnover is therefore due to slowly-rotating
black holes becoming radio-loud AGNs at a later cosmic epoch and one
can speculate on plausible scenarios for this. For example, a major
galaxy merger can produce a rapidly-spinning black hole (e.g., Wilson
\& Colbert 1995) as well as promoting vigorous star formation that
exhausts the cold gas supply, leading to a luminous radio mode
AGN. Galaxies that do not undergo such mergers will not have their
black holes spun up and will retain their cold gas for longer,
allowing them to operate in quasar-mode until a later cosmic epoch
when they eventually become low-luminosity radio mode AGNs.

A major observational challenge that hampers further study of the
radio-mode AGN population's evolution is the difficulty in separating
these objects from the rapidly evolving quasar-mode sources, both
radio-loud and radio-quiet. If this is to be done by using the
infrared--radio correlation, as we have, it requires mid-infrared
imaging to levels $\ll1\,$mJy, which is beyond the capabilities of
\textit{Herschel\/} and hence is only possible in the immediate future
within the regions of sky that already possess deep MIPS
imaging. These cover just a few square degrees of the sky and so this
method is not suitable for the analysis of faint radio sources from
the large-area surveys about to be undertaken by new radio
telescopes. Section~\ref{sec:stellarpops} suggests that a method based
on the stellar population of the host galaxy might be viable, and this
is supported by the work of Best \& Heckman (2011).

\section{Summary}

We have presented spectra for 277 of the 505 radio sources in the
sample of Simpson et al.\ (2006), with reliable redshifts for 256, and
a further 11 possible redshifts. Spectroscopic observations of a
further 129 sources failed to produce any significant continuum or
line detections.  Photometric redshifts have been derived for the
entire sample, bar seven sources strongly affected by foreground
objects.

We have found the observed redshift distribution of our sources to be
consistent with that predicted by recent semi-empirical simulations,
and have found that the median $K$-band magnitude of radio sources
does not change with radio flux density. However, we have demonstrated
that sources with $S_{\rm1.4GHz}\la1$\,mJy are systematically offset
from the $K$--$z$ relation to fainter magnitudes, compared to sources
at brighter radio fluxes. This argues strongly against the use of
`single-band photometric redshifts' for such sources.

We have used the infrared--radio correlation to separate our sample
into radio-loud and radio-quiet sources and investigate the stellar
populations of both types. We found that almost all radio-quiet
sources show evidence for recent activity, while the stellar
properties of radio-loud AGNs vary as a function of radio luminosity,
with the fraction of galaxies showing recent activity increasing from
almost zero at $L_{\rm1.4GHz}\approx10^{23}\rm\,W\,Hz^{-1}$ to nearly
100\,per cent at $L_{\rm1.4GHz}>10^{26}\rm\,W\,Hz^{-1}$. Even among
the predominantly old host galaxies, we find few sources that satisfy
the `pBzK' criterion of Daddi et al.\ (2004). 

We have calculated the radio luminosity function in redshift bins out
to $z=4$ and found more rapid evolution among the most luminous
sources. However, we find that the evolution at intermediate radio
luminosities is a combination of strong evolution among the radio-quiet
population and very weak evolution among radio-loud AGNs. We find
hints of a decline in the space density of radio-loud sources at
$z\ga1$, with the redshift of the turnover being dependent on radio
luminosity. While qualitatively consistent with other studies, we find
an earlier turnover than has been found previously. Our sample is
large enough to show evidence for a turnover at luminosities well
below the break in the local radio luminosity function, where the
radio-loud population is dominated by sources undergoing
radiatively-inefficient accretion that, in the current paradigm,
curtails the cooling of halo gas and hence inhibits further star
formation. This is different from the `downsizing' seen in the X-ray
AGN population as the relationship between radio luminosity and black
hole mass is much weaker at these luminosities. Instead, if radio
luminosity is primarily dependent on black hole spin, this suggests
that the fuel supply for slowly-rotating black holes is exhausted
later than for rapidly-rotating holes.

Further analysis of this effect is hampered by the strong evolution
within the radio-quiet population, which therefore contributes
significantly to the source counts at these luminosities. Although we
have used mid-infrared data to classify objects as either radio-loud
or radio-quiet, similar data exist over only small regions of the sky
and is is necessary to investigate alternative classification methods
in order to take full advantage of the large samples of these objects
about to be discovered by new radio telescopes. New radio data in the
UDS reach an r.m.s.\ sensitivity of $\sim6\,\mu$Jy\,beam$^{-1}$ and
provide a sample of over 1500 radio sources with deep optical and
near-infrared imaging, including deeper imaging in the $z'$ and $u^*$
filters (the latter to be described in Foucaud et al., in preparation)
and new $Y$-band images from the VISTA VIDEO Public Survey. These data
will improve the photometric redshift estimates at $z\ga1$, and will
be complemented with the extensive spectroscopic observations from the
UDSz program (Almaini et al., in preparation). Multi-colour
\textit{Hubble Space Telescope\/} imaging has also recently been taken
as part of the Cosmic Assembly Near-Infrared Deep Extragalactic Legacy
Survey (CANDELS; Grogin et al., in preparation), allowing detailed
galaxy morphologies to be studied over approximately 10\,per cent of
the combined SXDF/UDS field.

\section*{Acknowledgments}

This paper is based in large part on data collected at the European
Organisation for Astronomical Research in the Southern Hemisphere,
Chile as part of programme 074.A-0333, but also makes use of
additional data obtained at other telescopes. The Australian
Astronomical Observatory is a division of the Department of
Innovation, Industry, Science and Research and, at the time of our
observations, was also funded in part by the United Kingdom Particle
Physics and Astronomy Research Council.  The William Herschel
Telescope is operated on the island of La Palma by the Isaac Newton
Group in the Spanish Observatorio del Roque de los Muchachos of the
Instituto de Astrof\'{\i}sica de Canarias.  The Gemini Observatory is
operated by the Association of Universities for Research in Astronomy,
Inc., under a cooperative agreement with the NSF on behalf of the
Gemini partnership: the National Science Foundation (United States),
the Science and Technology Facilities Council (United Kingdom), the
National Research Council (Canada), CONICYT (Chile), the Australian
Research Council (Australia), Minist\'{e}rio da Ci\^{e}ncia e
Tecnologia (Brazil) and Ministerio de Ciencia, Tecnolog\'{\i}a e
Innovaci\'{o}n Productiva (Argentina). The W.~M.~Keck Observatory is
operated as a scientific partnership among the California Institute of
Technology, the University of California and the National Aeronautics
and Space Administration. The Observatory was made possible by the
generous financial support of the W.~M.~Keck Foundation. The United
Kingdom Infrared Telescope is operated by the Joint Astronomy Centre
on behalf of STFC. Subaru Telescope is operated by the National
Astronomical Observatory of Japan. The authors wish to recognize and
acknowledge the very significant cultural role and reverence that the
summit of Mauna Kea has always had within the indigenous Hawaiian
community. We are most fortunate to have the opportunity to conduct
observations from this mountain. This work is also based in part on
observations made with the \textit{Spitzer Space Telescope\/}, which
is operated by the Jet Propulsion Laboratory, California Institute of
Technology under a contract with NASA. It also makes use of data
products from the Two Micron All Sky Survey, which is a joint project
of the University of Massachusetts and the Infrared Processing and
Analysis Center/California Institute of Technology, funded by the
National Aeronautics and Space Administration and the National Science
Foundation.

The authors thank the Science and Technology Facilities Council for
funding and are extremely grateful to Carlo Izzo for his help with the
VIMOS data reduction. We thank Philip Best for commenting on a draft
of this manuscript and Ian Smail for making his reduced spectra
available in advance of publication. We appreciate the work of the
members of the UKIDSS UDS Working Group and the UDSz project,
in particular Michele Cirasuolo, Ross McLure, and Henry Pearce.

\appendix
\section{Optical spectra of radio sources}

This figure is available in the online journal.

\begin{figure*}
\resizebox{\colwidth}{!}{\includegraphics[angle=-90]{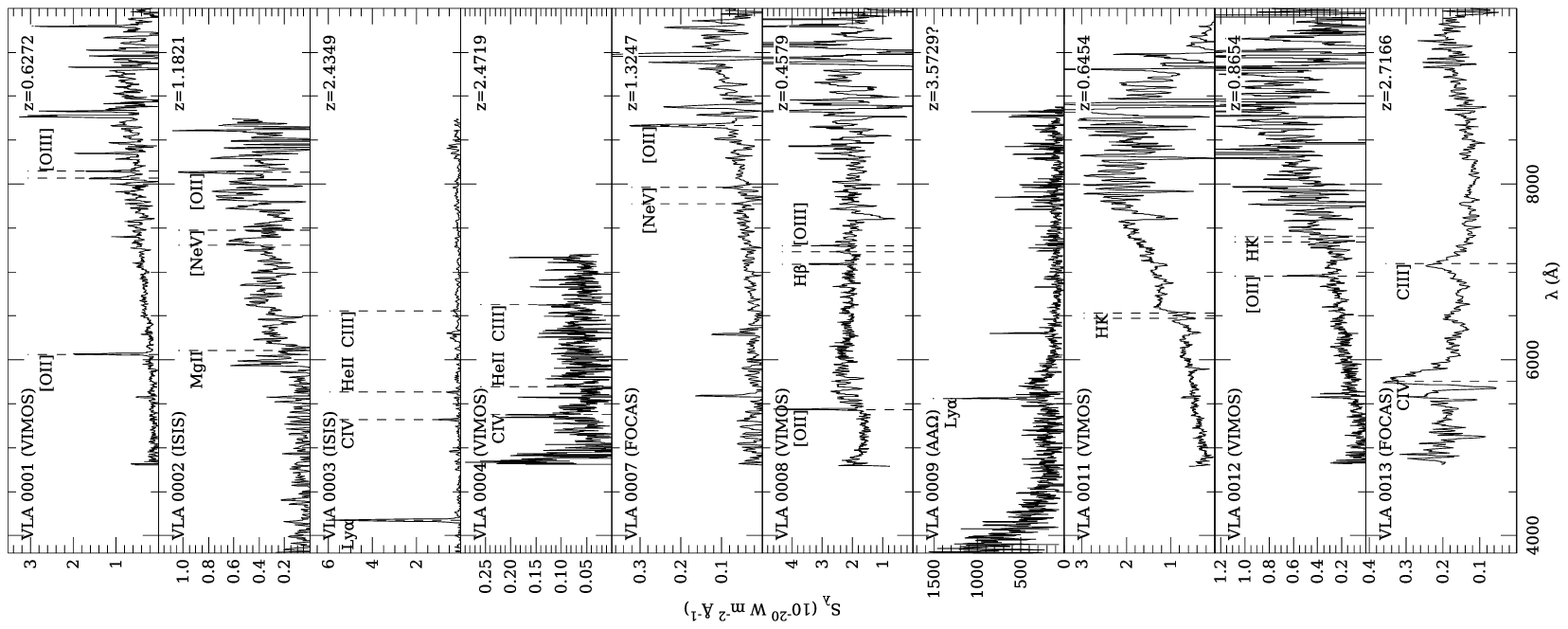}}
\resizebox{\colwidth}{!}{\includegraphics[angle=-90]{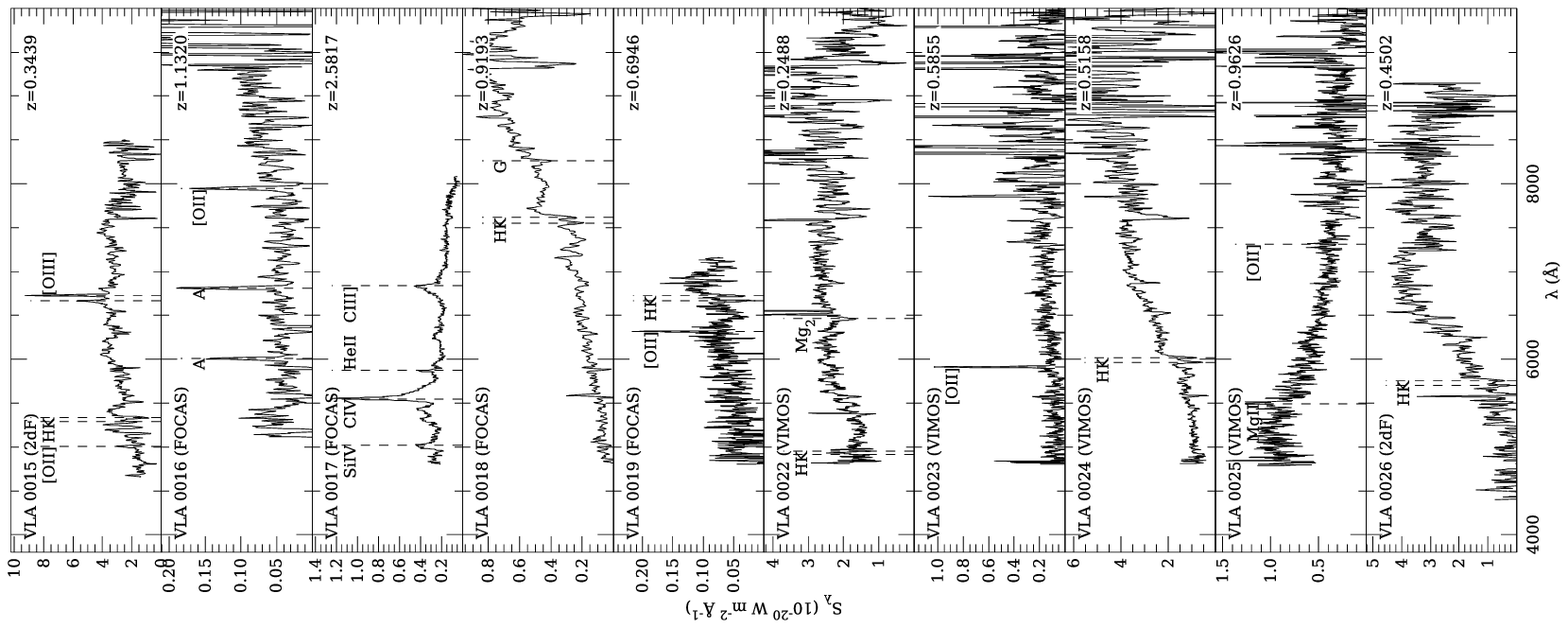}}
\caption[]{Spectroscopy of SXDF radio sources. The names of the
  sources and the origin of the spectra are shown in the top left
  corner of each panel, while the redshift is displayed in the top
  right. Identified features are marked with vertical dashed lines and
  labelled. An `A' indicates a feature is an artifact. Some spectra
  have been lightly smoothed or had strong artifacts interpolated
  across for presentation purposes.}
\end{figure*}

\addtocounter{figure}{-1}
\begin{figure*}
\resizebox{\colwidth}{!}{\includegraphics[angle=-90]{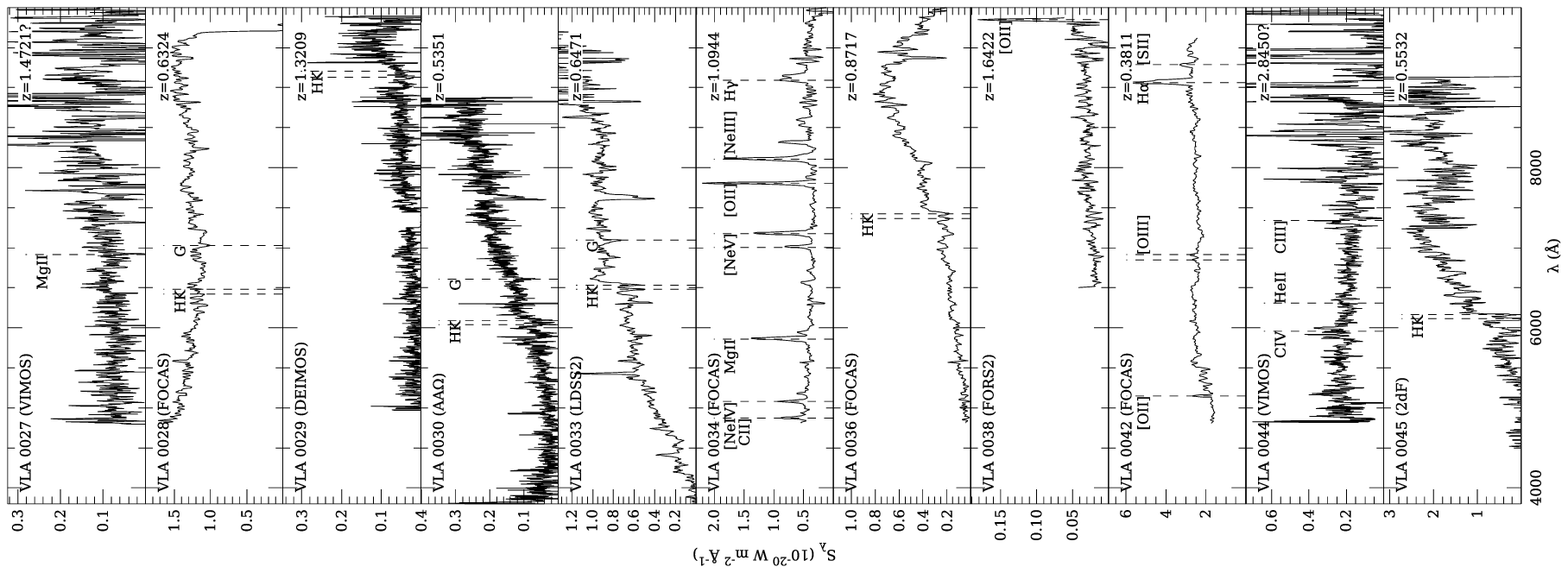}}
\resizebox{\colwidth}{!}{\includegraphics[angle=-90]{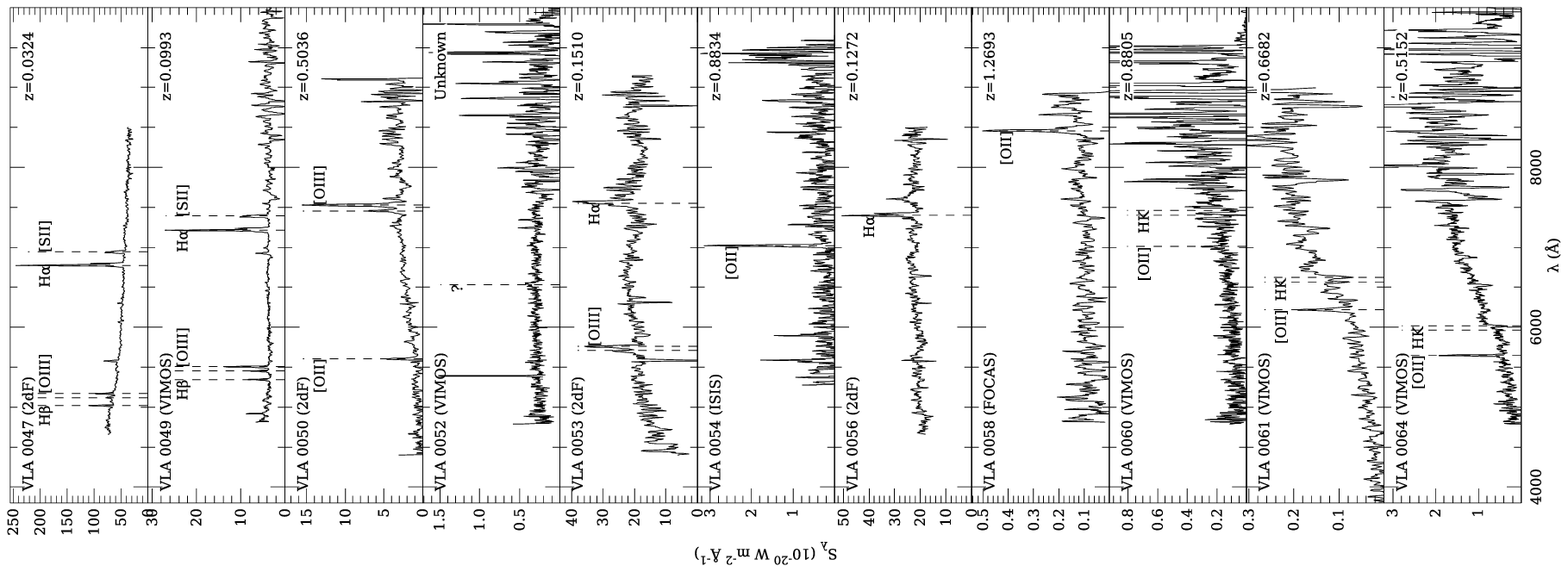}}
\caption[]{\textit{continued}.}
\label{fig:spectra}
\end{figure*}

\addtocounter{figure}{-1}
\begin{figure*}
\resizebox{\colwidth}{!}{\includegraphics[angle=-90]{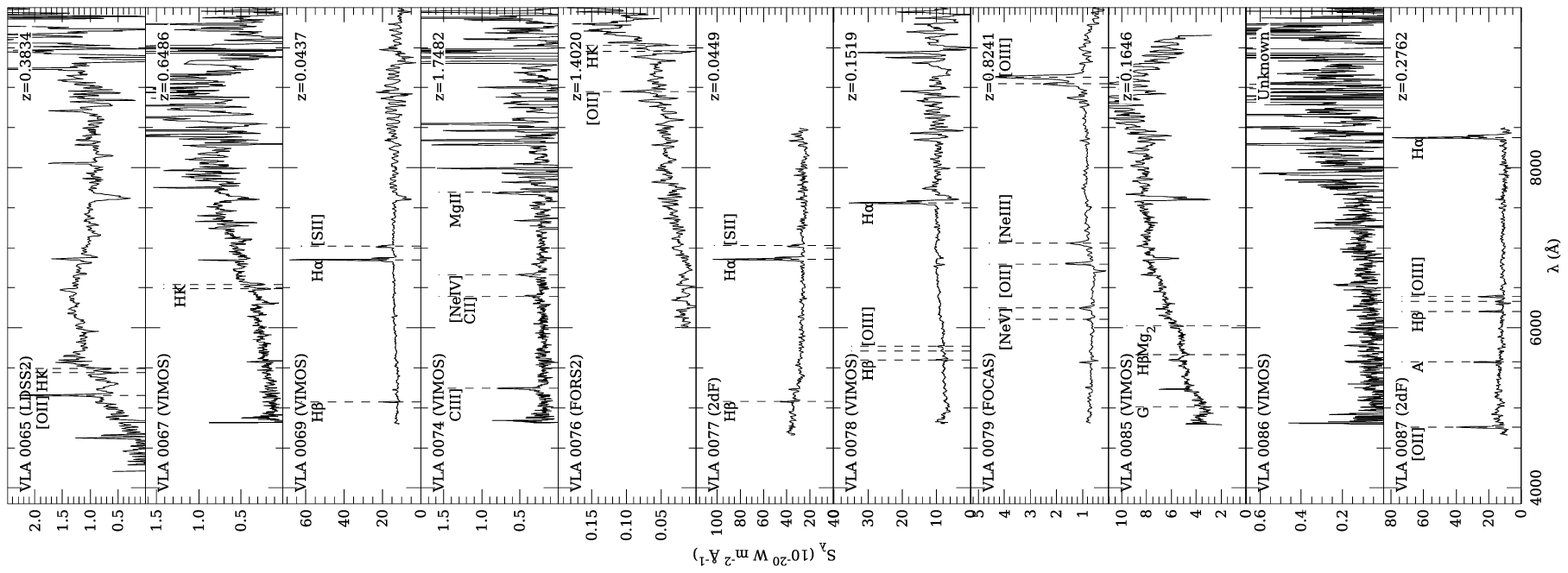}}
\resizebox{\colwidth}{!}{\includegraphics[angle=-90]{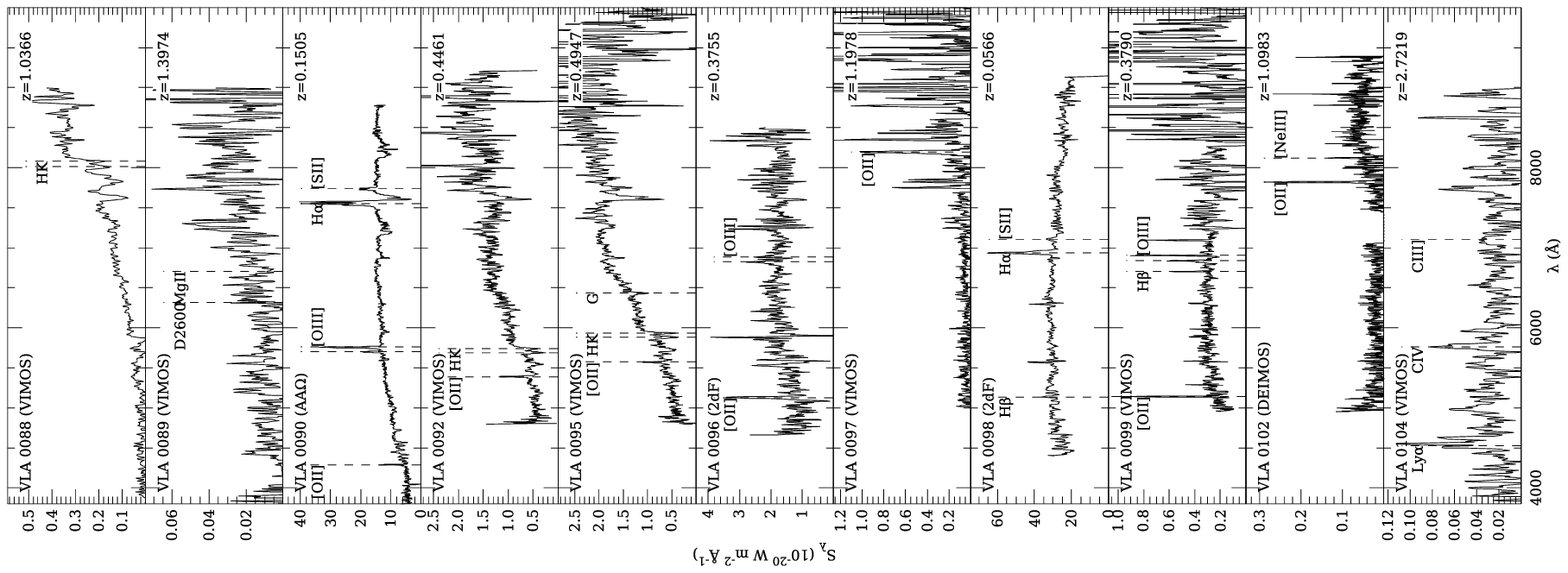}}
\caption[]{\textit{continued}.}
\end{figure*}

\addtocounter{figure}{-1}
\begin{figure*}
\resizebox{\colwidth}{!}{\includegraphics[angle=-90]{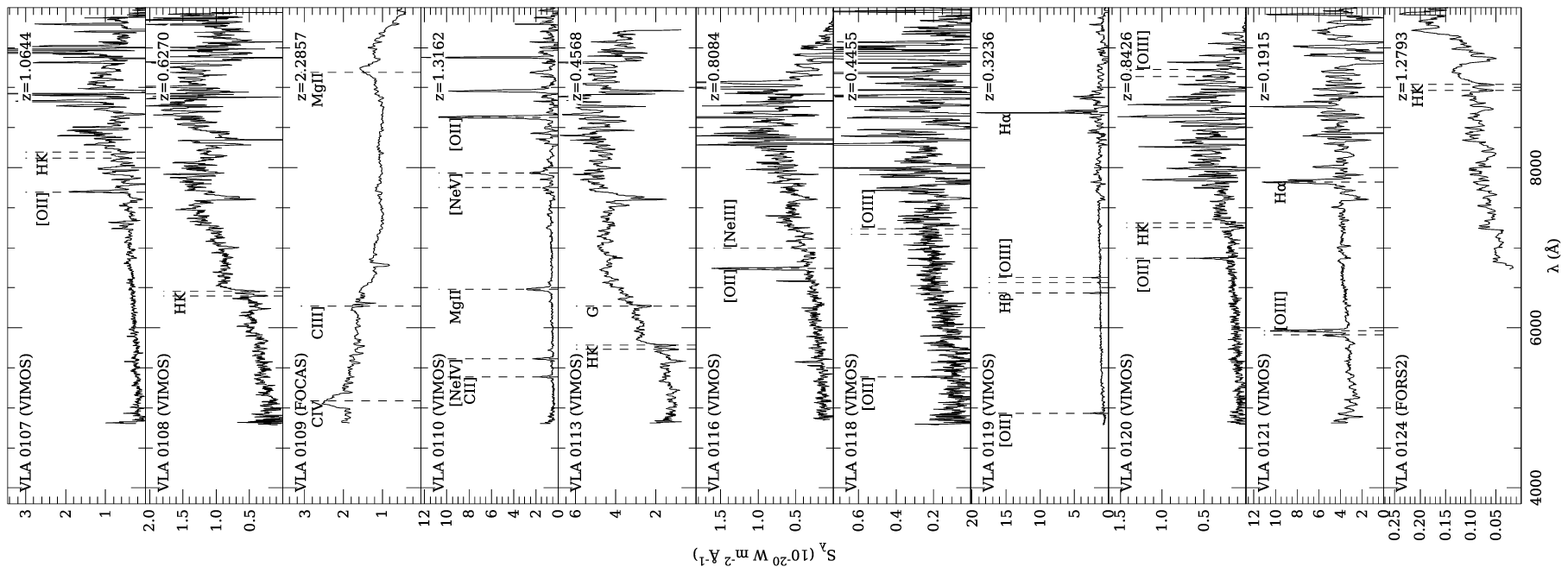}}
\resizebox{\colwidth}{!}{\includegraphics[angle=-90]{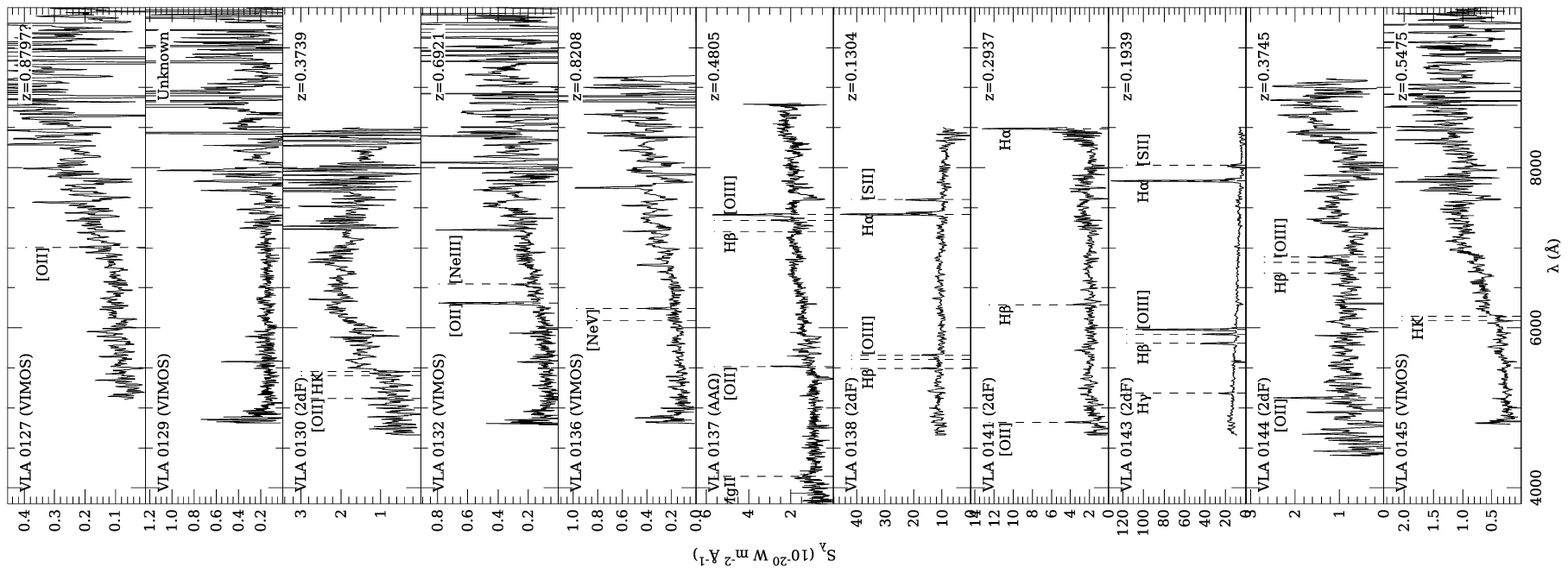}}
\caption[]{\textit{continued}.}
\end{figure*}

\addtocounter{figure}{-1}
\begin{figure*}
\resizebox{\colwidth}{!}{\includegraphics[angle=-90]{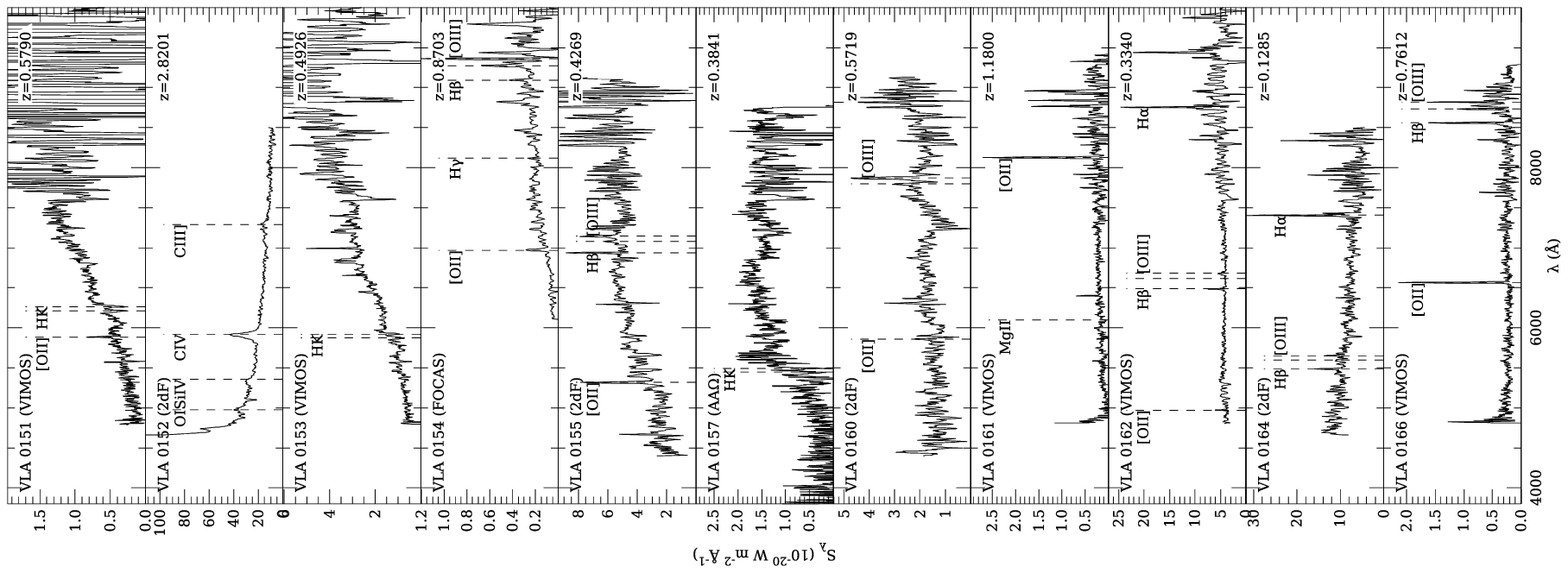}}
\resizebox{\colwidth}{!}{\includegraphics[angle=-90]{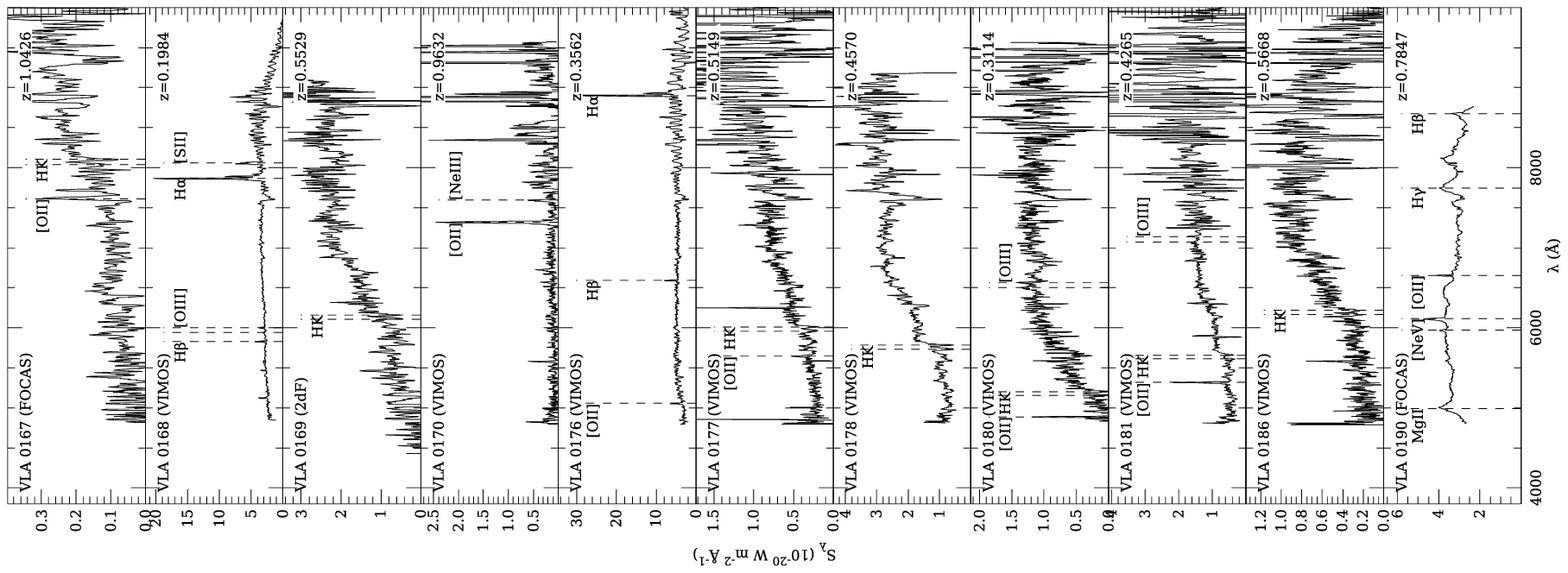}}
\caption[]{\textit{continued}.}
\end{figure*}

\addtocounter{figure}{-1}
\begin{figure*}
\resizebox{\colwidth}{!}{\includegraphics[angle=-90]{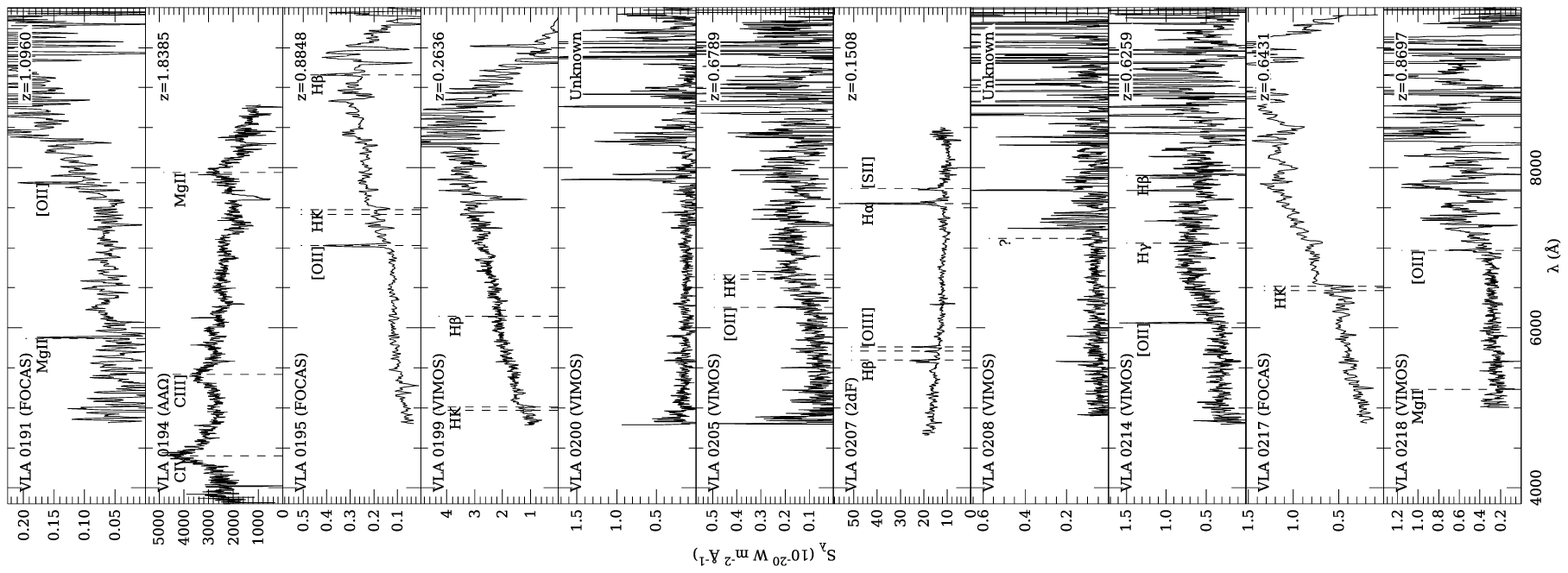}}
\resizebox{\colwidth}{!}{\includegraphics[angle=-90]{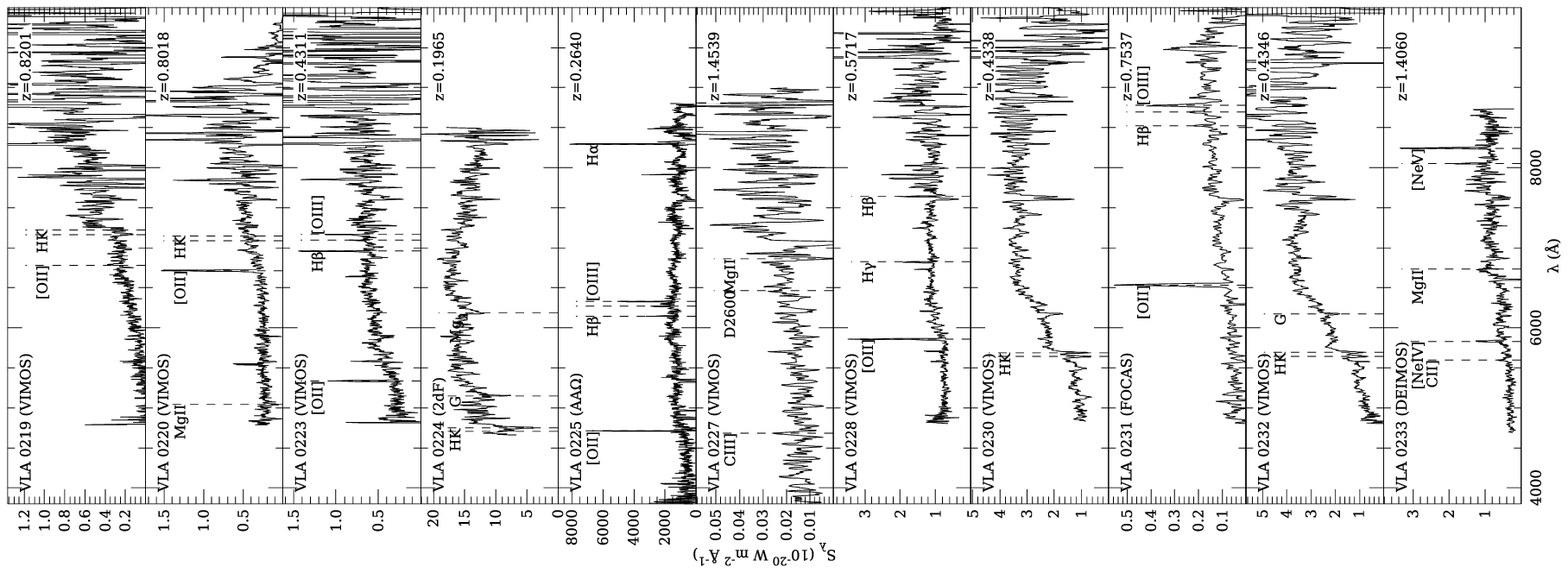}}
\caption[]{\textit{continued}.}
\end{figure*}

\addtocounter{figure}{-1}
\begin{figure*}
\resizebox{\colwidth}{!}{\includegraphics[angle=-90]{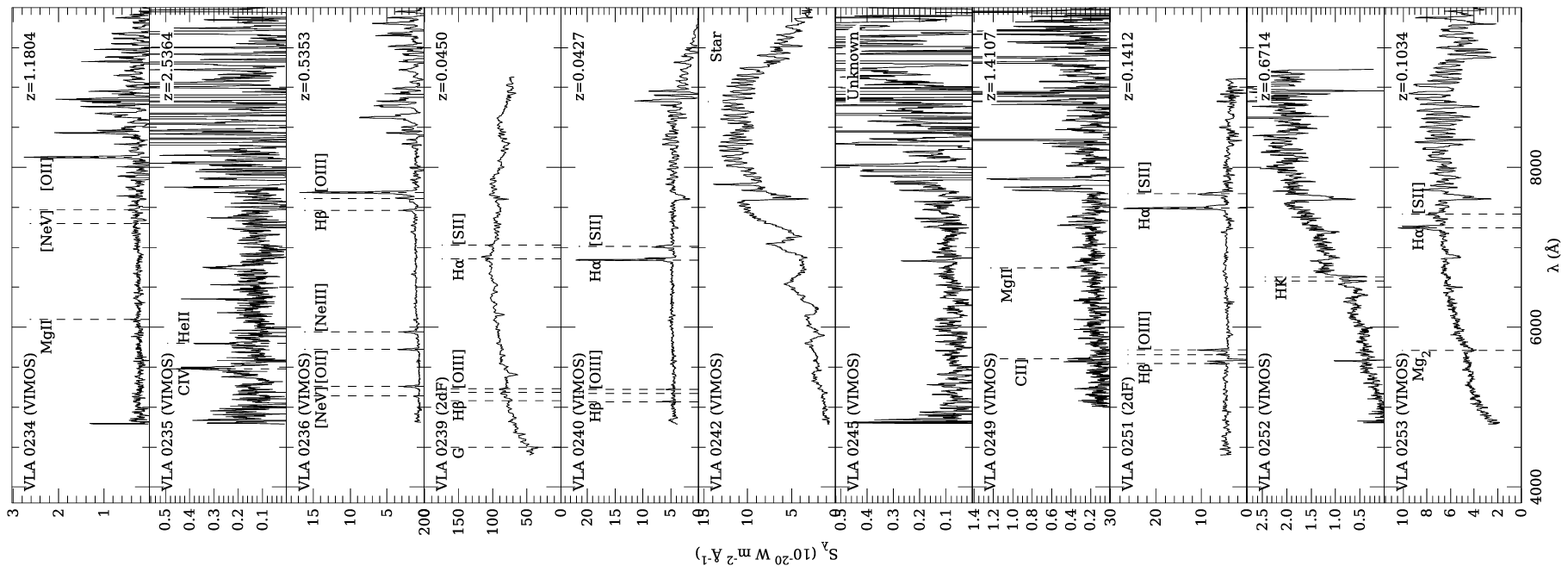}}
\resizebox{\colwidth}{!}{\includegraphics[angle=-90]{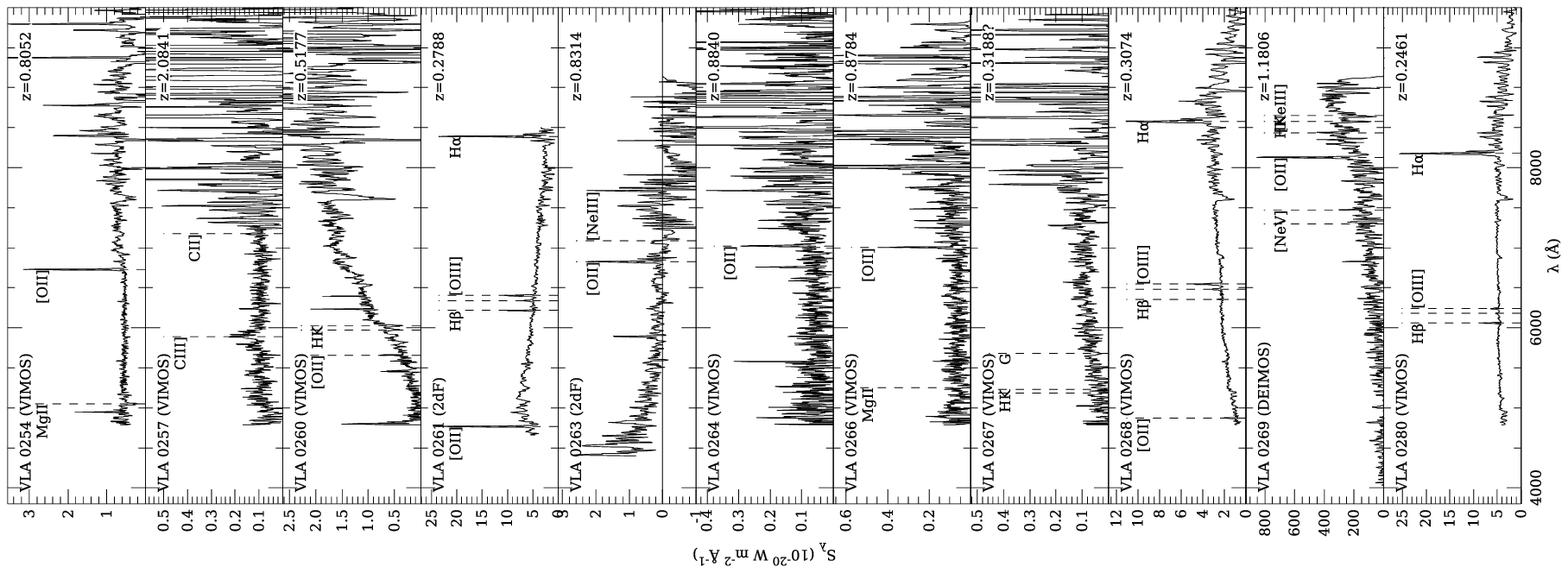}}
\caption[]{\textit{continued}.}
\end{figure*}

\addtocounter{figure}{-1}
\begin{figure*}
\resizebox{\colwidth}{!}{\includegraphics[angle=-90]{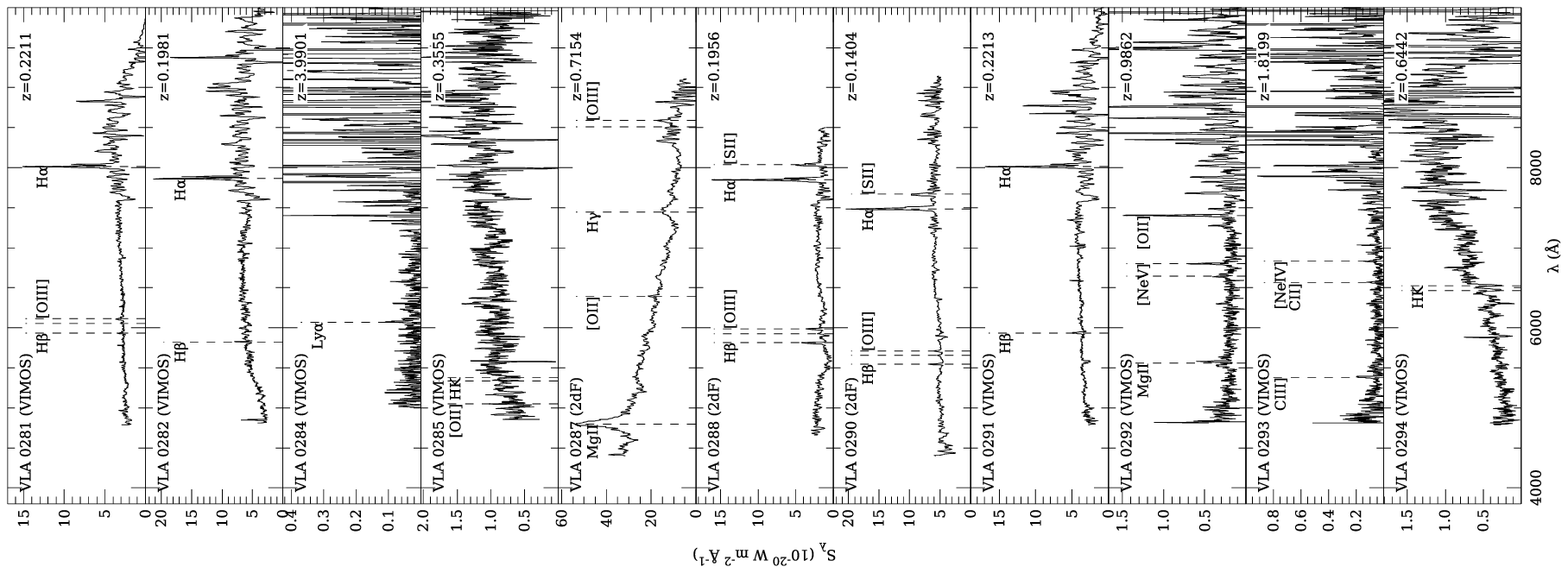}}
\resizebox{\colwidth}{!}{\includegraphics[angle=-90]{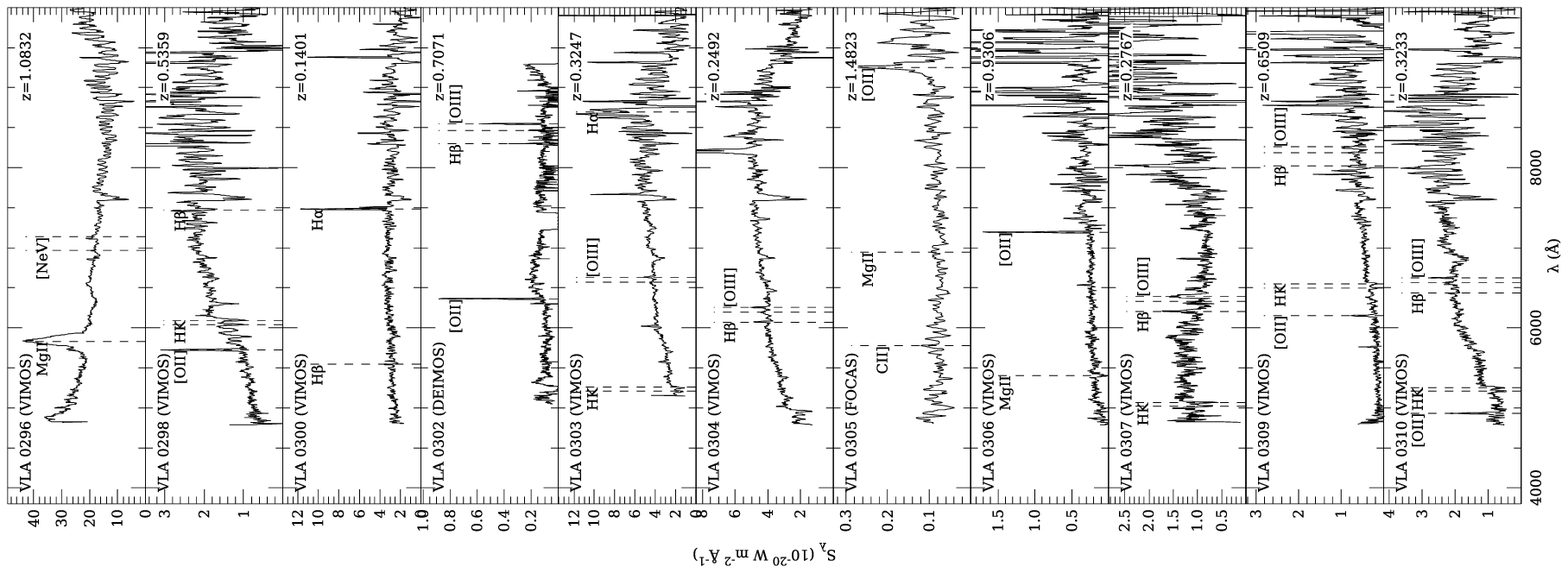}}
\caption[]{\textit{continued}.}
\end{figure*}

\addtocounter{figure}{-1}
\begin{figure*}
\resizebox{\colwidth}{!}{\includegraphics[angle=-90]{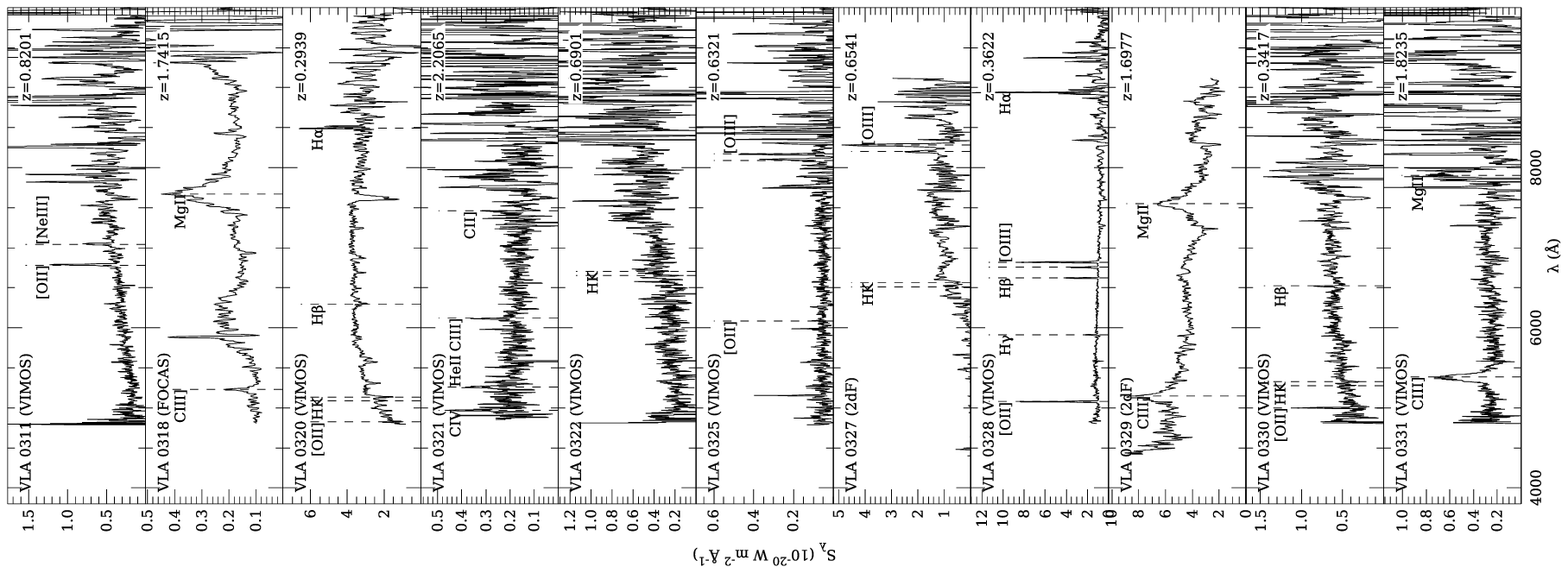}}
\resizebox{\colwidth}{!}{\includegraphics[angle=-90]{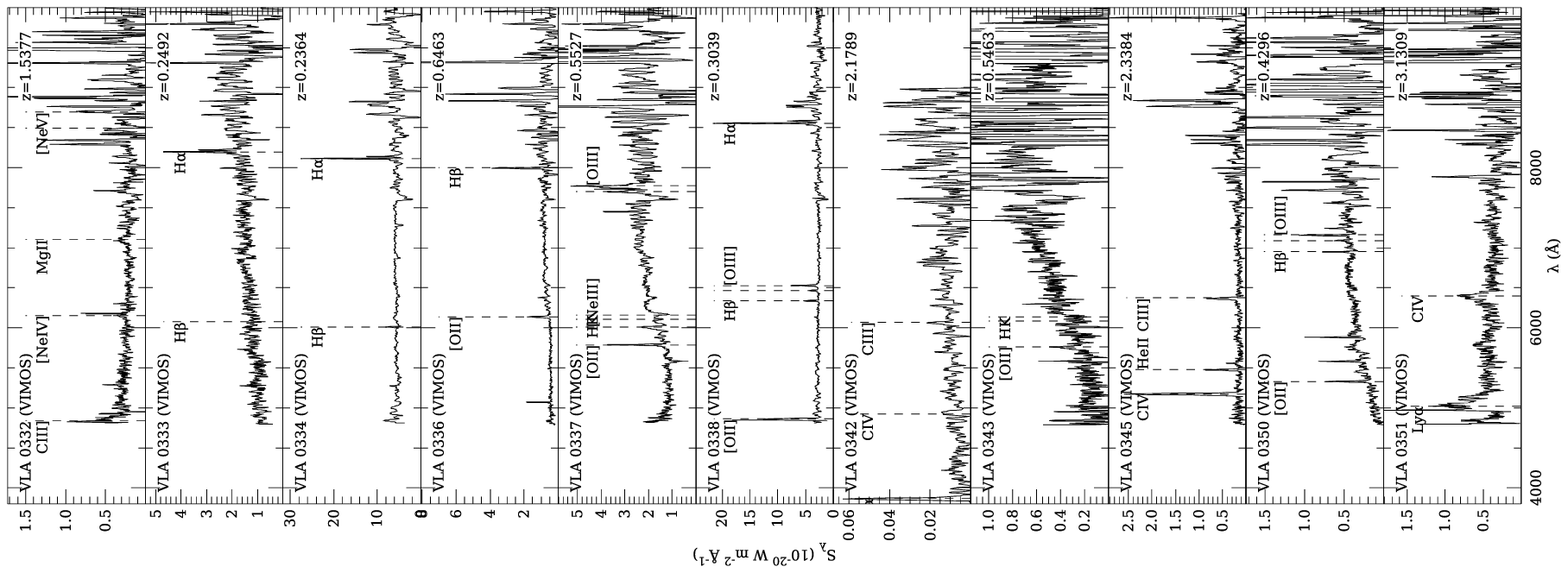}}
\caption[]{\textit{continued}.}
\end{figure*}

\addtocounter{figure}{-1}
\begin{figure*}
\resizebox{\colwidth}{!}{\includegraphics[angle=-90]{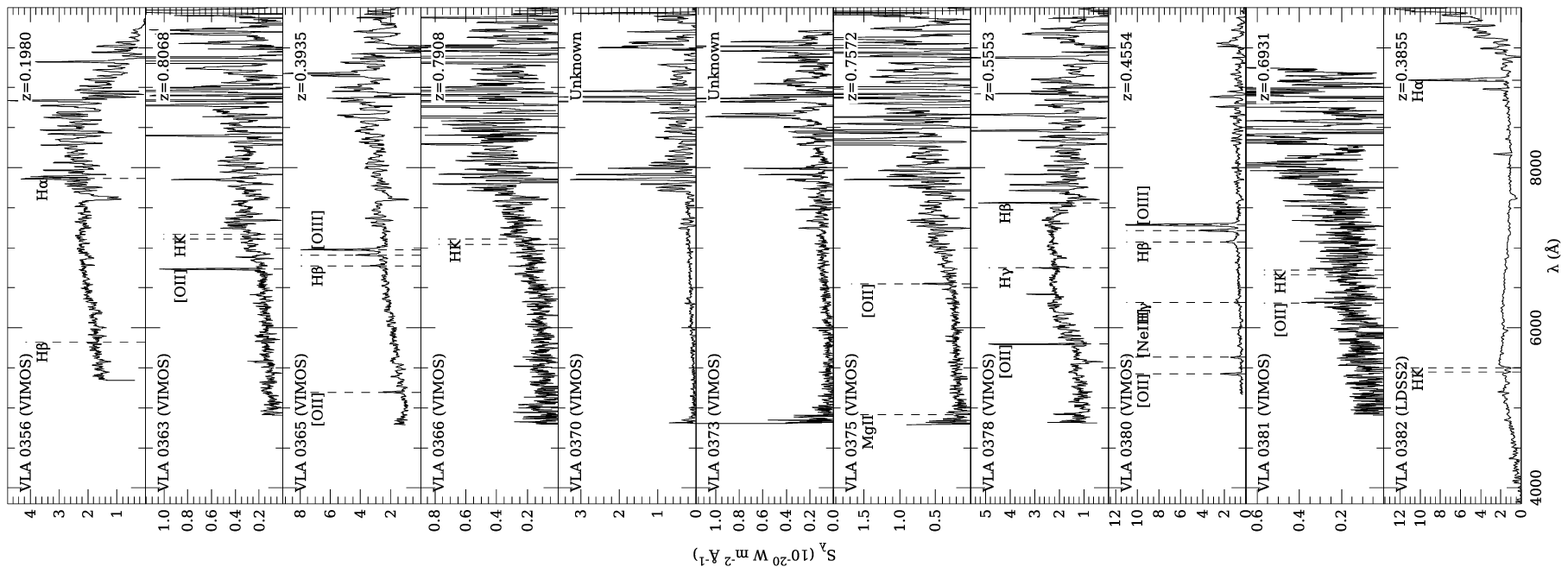}}
\resizebox{\colwidth}{!}{\includegraphics[angle=-90]{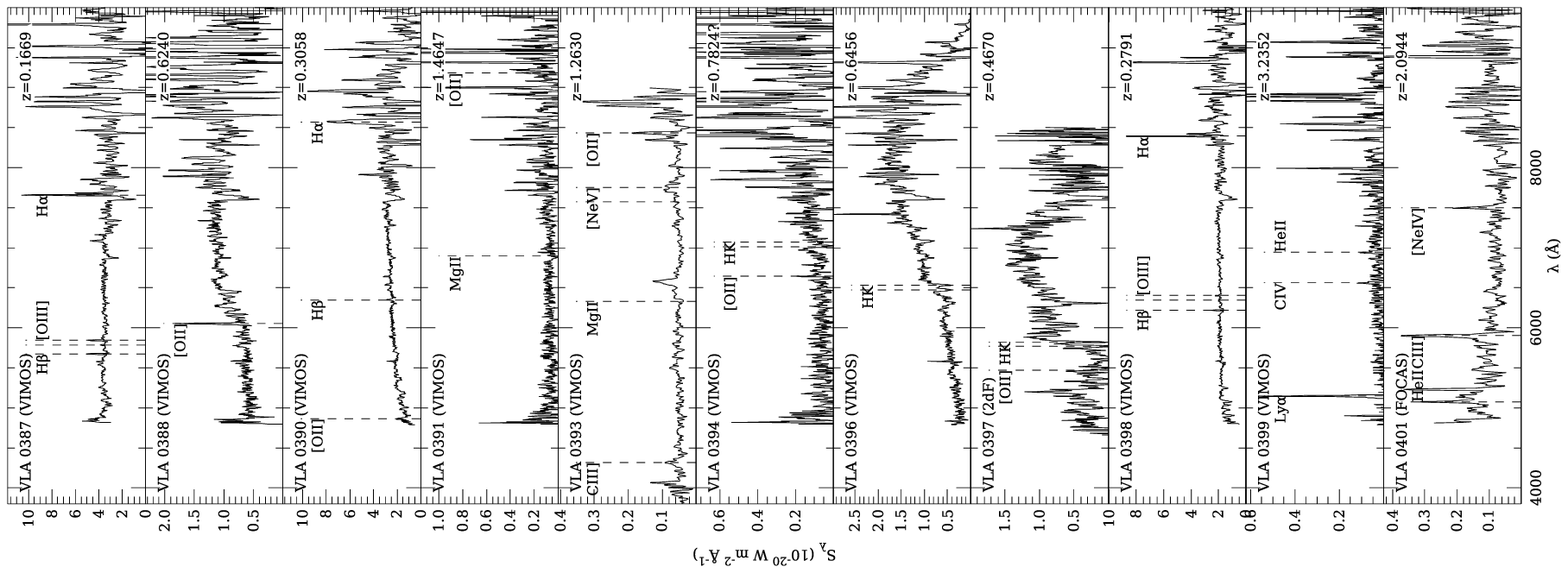}}
\caption[]{\textit{continued}.}
\end{figure*}

\addtocounter{figure}{-1}
\begin{figure*}
\resizebox{\colwidth}{!}{\includegraphics[angle=-90]{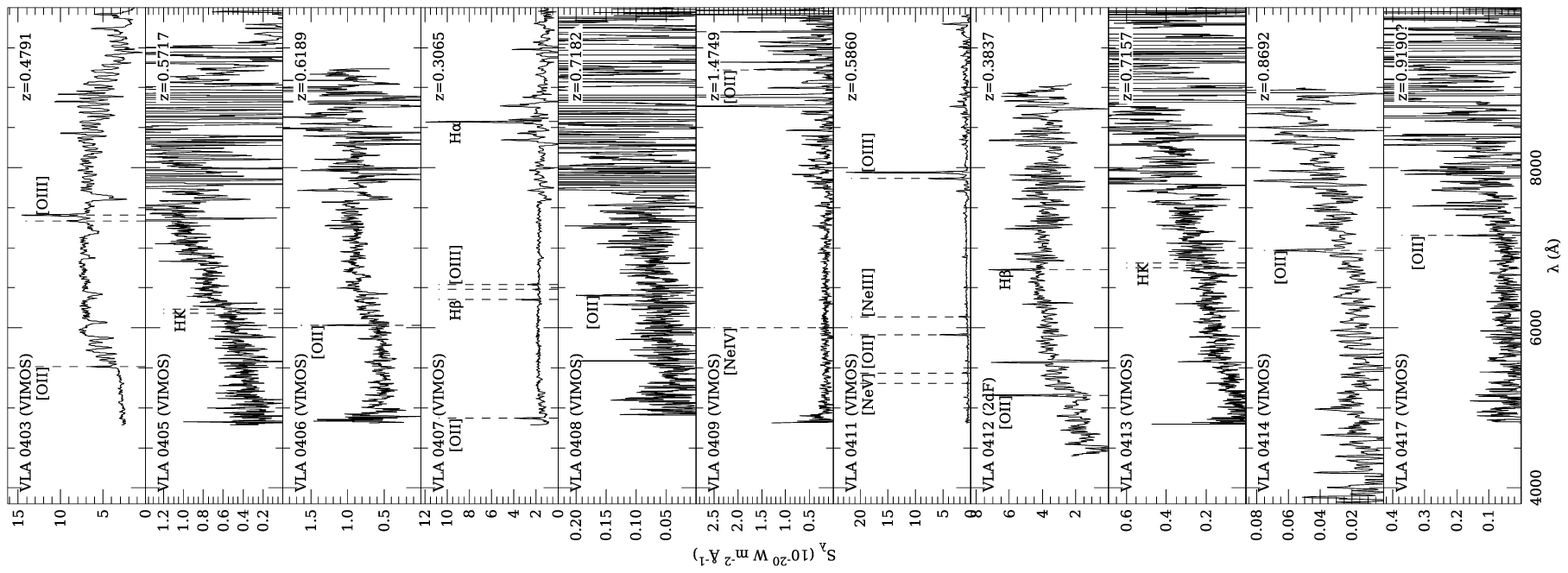}}
\resizebox{\colwidth}{!}{\includegraphics[angle=-90]{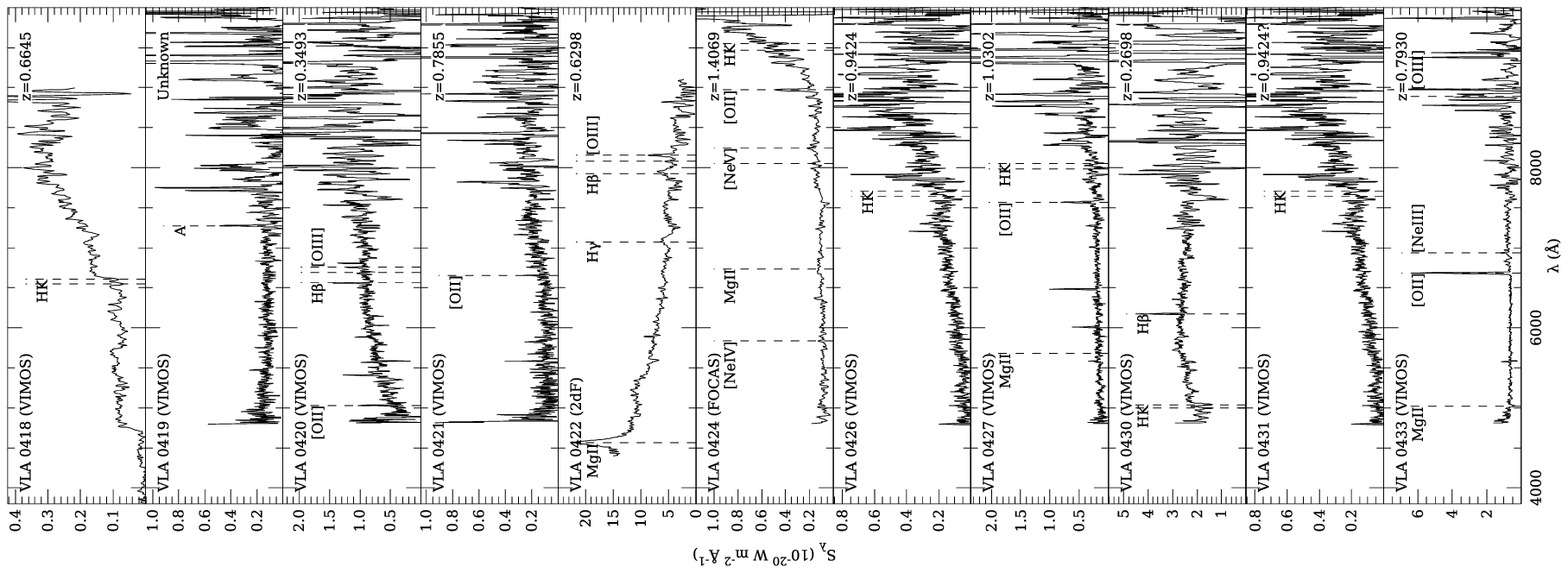}}
\caption[]{\textit{continued}.}
\end{figure*}

\addtocounter{figure}{-1}
\begin{figure*}
\resizebox{\colwidth}{!}{\includegraphics[angle=-90]{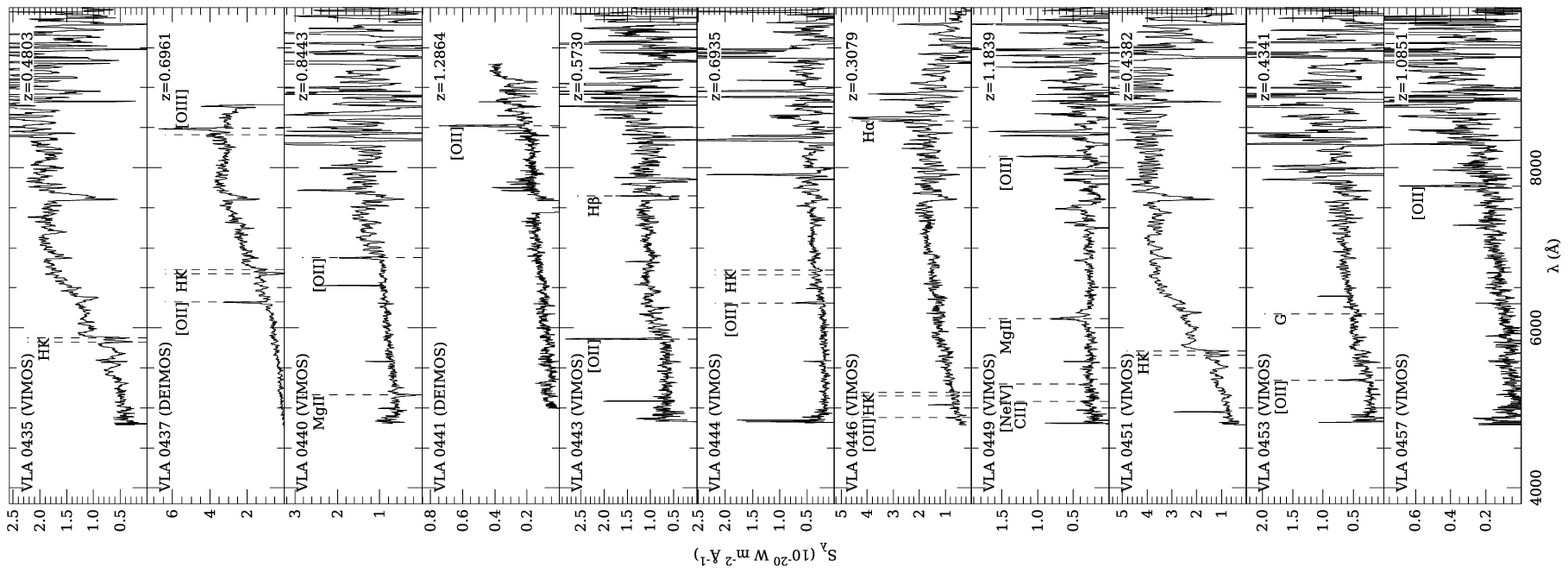}}
\resizebox{\colwidth}{!}{\includegraphics[angle=-90]{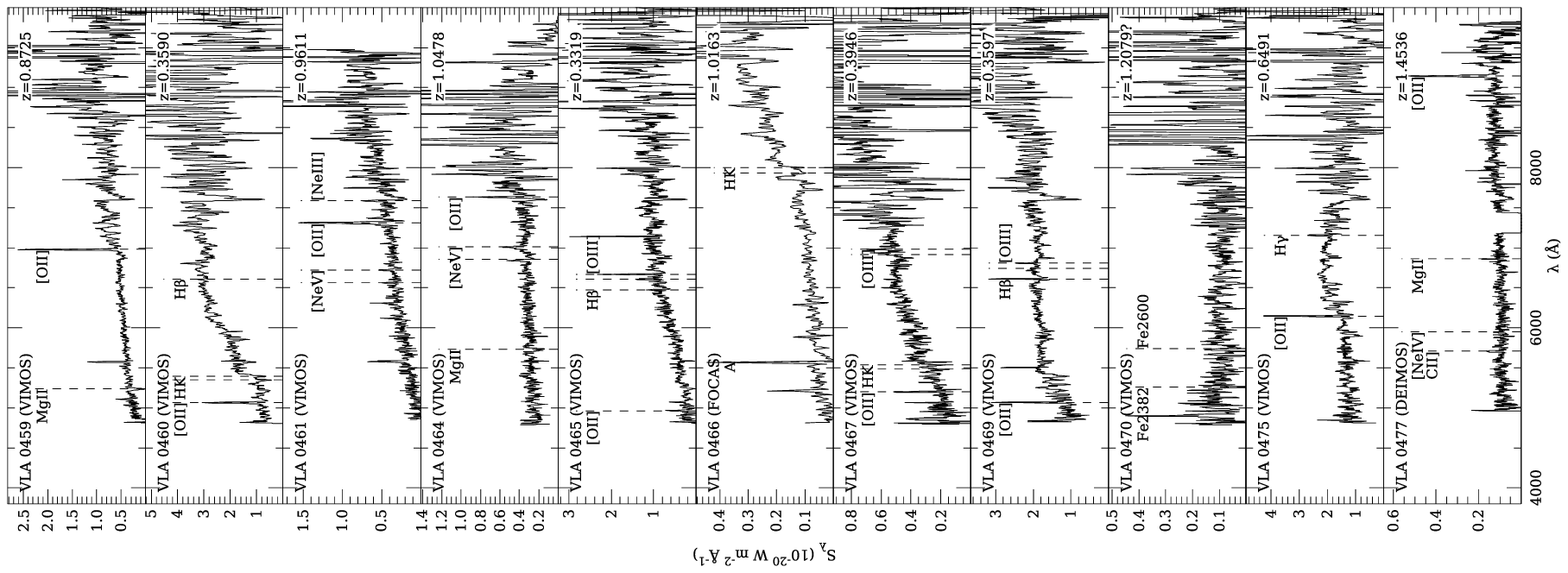}}
\caption[]{\textit{continued}.}
\end{figure*}

\addtocounter{figure}{-1}
\begin{figure*}
\resizebox{\colwidth}{!}{\includegraphics[angle=-90]{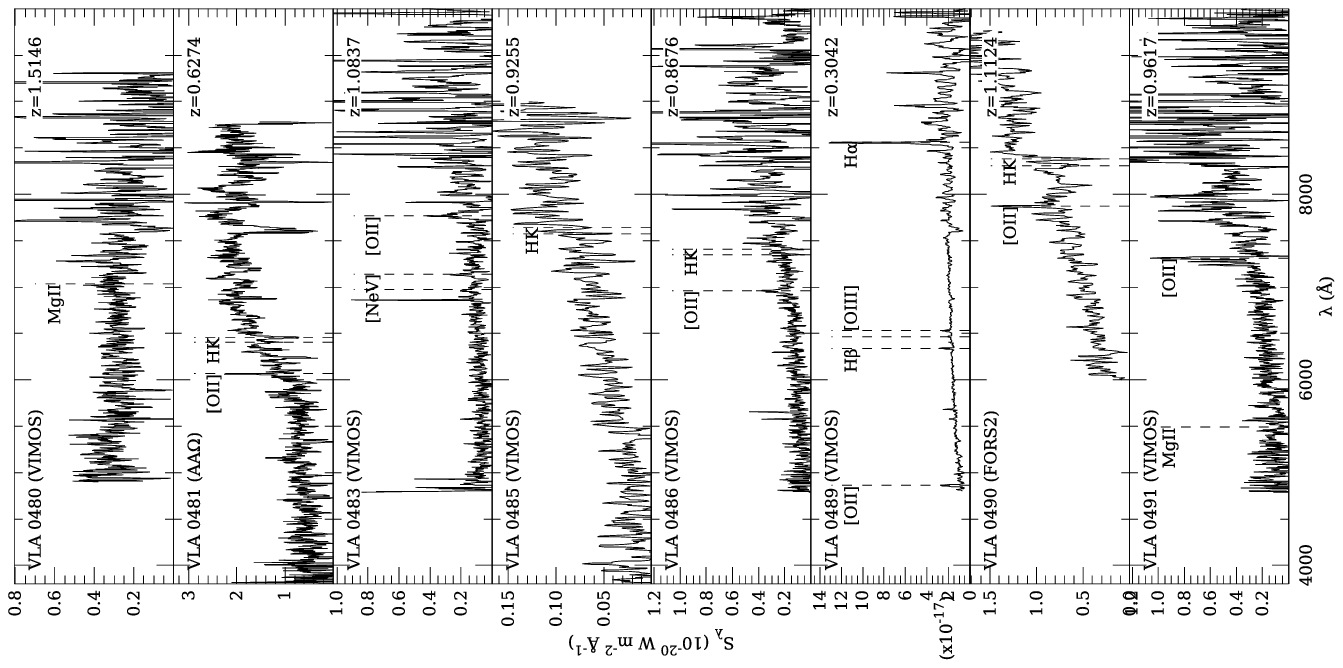}}
\resizebox{\colwidth}{!}{\includegraphics[angle=-90]{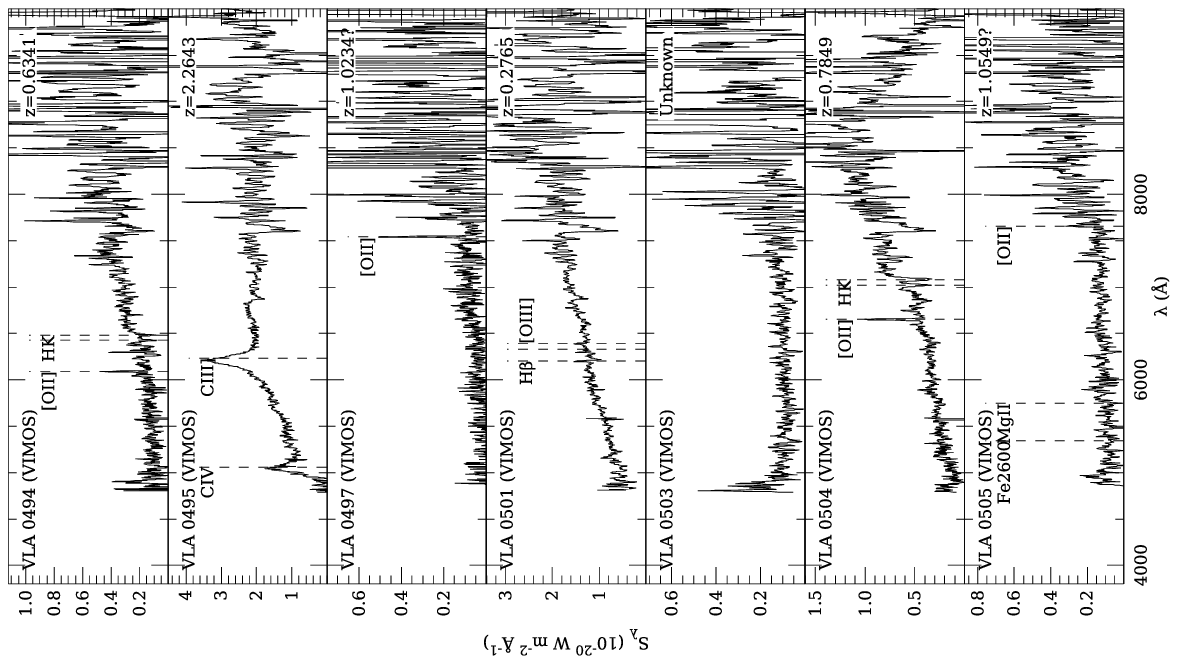}}
\caption[]{\textit{continued}.}
\end{figure*}

\label{lastpage}

\end{document}